\documentclass[sigconf]{acmart}
\setcopyright{none}
\renewcommand\footnotetextcopyrightpermission[1]{} % removes footnote with conference information in first column
\usepackage{array}

\makeatletter
\newcolumntype{"}{@{\hskip\tabcolsep\vrule width 2pt\hskip\tabcolsep}}
\makeatother

\usepackage{graphicx}
\usepackage{balance}  % for  \balance command ON LAST PAGE  (only there!)

\usepackage{url}
\usepackage[shortlabels]{enumitem}
\usepackage{algorithm}
\usepackage{amsmath}
\usepackage{wasysym}
\usepackage{textcomp}
\usepackage[T1]{fontenc}
\usepackage[noend]{algpseudocode}
\usepackage[caption=false,font=footnotesize]{subfig}
\usepackage[aboveskip=0pt]{caption}
\usepackage[belowskip=0pt]{caption}
\usepackage[compact]{titlesec}
\titlespacing{\section}{0pt}{1ex}{0ex}
\titlespacing{\subsection}{0pt}{1ex}{0ex}
\titlespacing{\subsubsection}{0pt}{0ex}{0ex}

\AtBeginDocument{%
  \providecommand\BibTeX{{%
    \normalfont B\kern-0.5em{\scshape i\kern-0.25em b}\kern-0.8em\TeX}}}

\begin{document}

%\title{Why Do My Blockchain Transactions Fail? A Study of Hyperledger Fabric [Experiment and Analysis]}
\title{Why Do My Blockchain Transactions Fail? A Study of Hyperledger Fabric (Extended version)*}
\thanks{*This is an extended version of an upcoming publication at ACM SIGMOD 2021. Please cite the original SIGMOD version.}

%\institution{Middleware Systems Research Group}
\author{Jeeta Ann Chacko}
\email{chacko@in.tum.de}
\affiliation{%
  \institution{Technical University of Munich}
}

\author{Ruben Mayer}
\email{mayerr@in.tum.de}
\affiliation{%
  \institution{Technical University of Munich}
}

\author{Hans-Arno Jacobsen}
\email{jacobsen@eecg.toronto.edu}
\affiliation{%
  \institution{University of Toronto}
}


\begin{abstract}
  Permissioned blockchain systems promise to provide both decentralized trust and privacy. Hyperledger Fabric is currently one of the most wide-spread permissioned blockchain systems and is heavily promoted both in industry and academia. Due to its optimistic concurrency model, the transaction failure rates in Fabric can become a bottleneck. While there is active research to reduce failures, there is a lack of understanding on their root cause and, consequently, a lack of guidelines on how to configure Fabric optimally for different scenarios. To close this gap, in this paper, we first introduce a formal definition of the different types of transaction failures in Fabric. Then, we develop a comprehensive testbed and benchmarking system, \textsc{HyperLedgerLab}, along with four different chaincodes that represent realistic use cases and a chaincode/workload generator. Using \textsc{HyperLedgerLab}, we conduct exhaustive experiments to analyze the impact of different parameters of Fabric such as block size, endorsement policies, and others, on transaction failures. We further analyze three recently proposed optimizations from the literature, Fabric++, Streamchain and FabricSharp, and evaluate under which conditions they reduce the failure rates. Finally, based on our results, we provide recommendations for Fabric practitioners on how to configure the system and also propose new research directions.
\end{abstract}

%\begin{CCSXML}
%<ccs2012>
%<concept>
%<concept_id>10002951.10002952</concept_id>
%concept_desc>Information systems~Data management systems</concept_desc>
%<concept_significance>500</concept_significance>
%/concept>
%</ccs2012>
%\end{CCSXML}

%\ccsdesc[500]{Information systems~Data management systems}


\maketitle
\fancyhead{}
%\clubpenalty = 0
%\widowpenalty = 0
%\displaywidowpenalty = 0
%\pagestyle{fancy}
%\fancyhf{}
%\balance
\section{Introduction}
\label{sec:introduction}

%With the advent of Bitcoin~\cite{article}, a renewed interest in decentralized trust and immutable records emerged. But enterprises are wary to adopt a framework which allows open participation and offers limited throughput. This gave birth to the concept of permissioned blockchains which provide higher transaction rates and decentralized trust, but restrict access to a set of authorized participants~\cite{brown2016corda, greenspan2015multichain, morgan2016quorum}. Permissioned blockchains are gaining increasing popularity since enterprises can now identify use cases which can be implemented more efficiently on blockchains. Following this trend, many permissioned blockchains such as Corda~\cite{brown2016corda}, Multichain~\cite{greenspan2015multichain} and Quorum~\cite{morgan2016quorum} appeared. Currently, Hyperledger Fabric (a.k.a. Fabric)~\cite{Androulaki:2018:HFD:3190508.3190538} is a widely used permissioned blockchain framework. For instance, a recent survey by Rauchs et al. shows that 48~\% of all live permissioned blockchain projects in the Cambridge Centre for Alternative Finance dataset build on Fabric~\cite{rauchs20192nd, ccaf}. 

With the advent of Bitcoin~\cite{article}, a renewed interest in decentralized trust and immutable records emerged. But enterprises are wary to adopt a framework which allows open participation and offers limited throughput. This gave birth to the concept of permissioned blockchains which provide higher transaction rates and decentralized trust~\cite{brown2016corda, greenspan2015multichain, morgan2016quorum}. As there exists a trade-off between decentralization, consistency and scalability in blockchains~\cite{8416397}, such permissioned blockchains need to restrict access to a set of authorized participants. Permissioned blockchains are gaining increasing popularity since enterprises can now identify use cases which can be implemented more efficiently on blockchains. Following this trend, many permissioned blockchains such as Corda~\cite{brown2016corda}, Multichain~\cite{greenspan2015multichain} and Quorum~\cite{morgan2016quorum} appeared. Currently, Hyperledger Fabric (a.k.a. Fabric)~\cite{Androulaki:2018:HFD:3190508.3190538} is a widely used permissioned blockchain framework. For instance, a recent survey by Rauchs et al. shows that 48\% of all live permissioned blockchain projects in the Cambridge Centre for Alternative Finance dataset build on Fabric~\cite{rauchs20192nd, ccaf}.

Despite its wide adoption, Fabric still has its limitations. It follows an optimistic concurrency control model which causes transaction failures when conflicting transactions are concurrently executed~\cite{HARDER1984111}. In our experiments, we observed that more than 40\% of the transactions failed due to concurrency-related conflicts in a realistic scenario (Electronic Health Record, cf. Table~\ref{cc}). There is active research to improve the throughput and reduce transaction failures in Fabric~\cite{8526892, Goel:2018:RFP:3284028.3284035, istvan2018streamchain, Sharma:2019:BLB:3299869.3319883, gorenflo2019xox, gorenflo2019fastfabric, 10.1145/3318464.3389693}. However, there is no formal definition for transaction failures, and no comprehensive study on their cause and the parameters that influence them. As a result, research on Fabric transaction failures falls short in exploring the full range of  the problem and trade-offs that are involved. For example, Fabric++ by Sharma et al.~\cite{Sharma:2019:BLB:3299869.3319883} reorders transactions to reduce a specific type of failures (Multi-Version Concurrency Control (MVCC) read conflicts), but neglects the effect on other failure types. Furthermore, different papers evaluate their approach to reduce concurrency-related transaction failures using distinct smart contracts and workloads, which hinders a direct and fair comparison~\cite{8526892, Goel:2018:RFP:3284028.3284035, istvan2018streamchain, Sharma:2019:BLB:3299869.3319883, gorenflo2019xox, gorenflo2019fastfabric, 10.1145/3318464.3389693}. For example, Sharma et al.~\cite{Sharma:2019:BLB:3299869.3319883} use a smart contract based on an asset transfer scenario while Istv{\'a}n et al.~\cite{istvan2018streamchain} use Fabcoin~\cite{Androulaki:2018:HFD:3190508.3190538}, a digital currency inspired by Bitcoin~\cite{article}.

\begin{figure*}
\centering
\includegraphics[width=\textwidth]{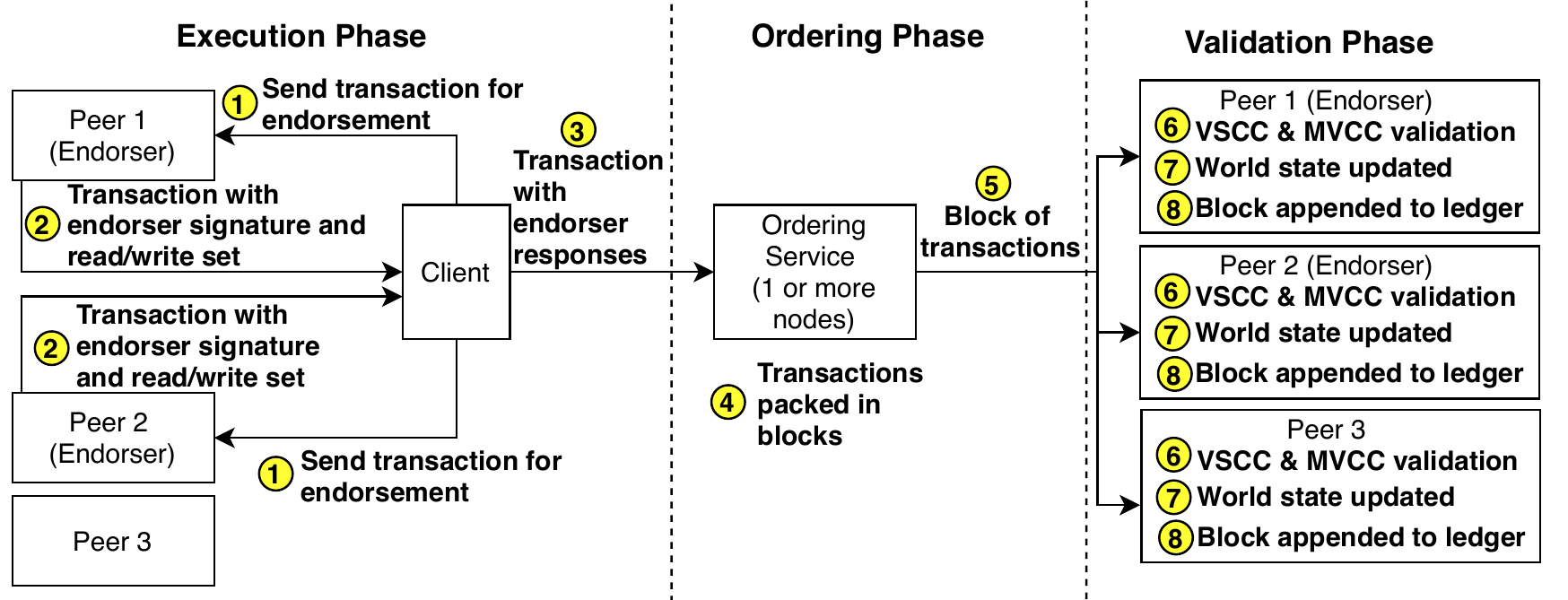}
\caption{Transaction Flow in Hyperledger Fabric}
\label{transactionflow}
\end{figure*}

In this paper, we close this gap by performing an extensive study on transaction failures in Fabric using a comprehensive benchmarking system. We provide the following contributions:

\begin{enumerate}[nosep]
	\item We formally define the different types of concurrency-related transaction failures in order to build a solid foundation for our study and for further research.
	\item We extensively study the various parameters influencing different transaction failures in Fabric as well as for three recent optimization techniques, Fabric++~\cite{Sharma:2019:BLB:3299869.3319883}, Streamchain~\cite{istvan2018streamchain} and FabricSharp~\cite{10.1145/3318464.3389693}. Our study reveals surprising insights and trade-offs regarding the configuration of Fabric. For instance, the \emph{block size} has a significant impact on the number of transaction failures at various transaction arrival rates---with the right block size, transaction failures could be reduced by up to 60\%.
	\item For realistic evaluations and controlled experiments, we develop a new testbed integrated with an extension of the Caliper benchmarking system, \textsc{HyperLedgerLab}~\cite{fablab}, which includes four new smart contracts that represent realistic use cases as well as a chaincode and workload generator. We released \textsc{HyperLedgerLab} as well as all the chaincodes and the generator as an open source project, so that other researchers can benefit from it and compare their work in a fair way. 
	%\footnote{Name changed to preserve double blind policy}
%	\item To facilitate our study, we develop a new testbed and benchmarking system, \textsc{HyperLedgerLab}, which includes four new smart contracts that represent realistic use cases as well as a chaincode/workload generator that can generate synthetic smart contracts and workloads. We plan to release \textsc{HyperLedgerLab} as an open source project, so that other researchers can benefit from it and compare their work in a fair way.
	\item  We identify best practices and principles for Fabric developers and also derive promising research directions that the scientific community can pursue in the future.
\end{enumerate}

The rest of this paper is organized as follows. In Section~\ref{sec:background}, we provide the technical background on Fabric. We introduce and formalize the different transaction failures in Section~\ref{sec:failures}. In Section~\ref{sec:benchmark}, we explain our new benchmarking system \textsc{HyperLedgerLab}, before we describe and discuss our experimental results in Section~\ref{sec:evaluation}. Finally, in Section~\ref{sec:discussion}, we summarize the insights and lessons learned, discuss related work in Section~\ref{sec:related} and conclude the paper in Section~\ref{sec:conclusion}.

\balance
\section{Hyperledger Fabric}
\label{sec:background}

Fabric is a popular open source permissioned blockchain system established under the Linux foundation~\cite{Androulaki:2018:HFD:3190508.3190538}. It is the first blockchain system that supports the creation of smart contracts in general purpose languages. Fabric allows clients to submit \emph{transactions} to a blockchain system which offers decentralized control of a shared, distributed state, i.e., there is not a single trusted entity that decides about the current state of the system. All possible functions that can be invoked by a transaction are defined in a \emph{smart contract}, which is called \emph{chaincode} in Fabric jargon. The distributed state, called \emph{world state}, is maintained as a versioned key-value store---Fabric currently supports LevelDB ~\cite{leveldb} and CouchDB ~\cite{couchdb}. Each key has a version number which is updated with every write. The \emph{distributed ledger} maintains the complete history of all the transactions (successful and failed) in the network which are grouped into blocks.  Both the world state and the distributed ledger are replicated on a set of distributed nodes that are registered on the Fabric network---called \emph{peers}. Peers receive blocks of transactions from an \emph{ordering service} that guarantees the ordered delivery; more precisely, all peers receive all transactions in the same order. They validate every transaction independently and update their copy of the world state and the ledger accordingly. For fault tolerance, the ordering service can be replicated and uses a consensus protocol (e.g., Paxos~\cite{lamport2001paxos} or Raft~\cite{Ongaro:2014:SUC:2643634.2643666}) to reach an agreement about the order of all transactions. The ordering service handles multiple \emph{channels}, i.e., private communication routes between different Fabric components. \emph{Endorsers} are a subset of the peers that have the additional role of endorsing transactions submitted by the clients, i.e., they simulate the execution of the transaction to generate \emph{read sets} and \emph{write sets} based on the current world state. An \emph{endorsement policy} defines the number of endorsements that are required for a transaction to be accepted as valid. Finally, peers are grouped into \emph{organizations} which typically correspond to real organizations or branches of an enterprise; these organizations can play an important role in the endorsement policy.

The transaction flow in Fabric follows three phases: execution, ordering and validation. This is referred to as the Execute-Order-Validate (E-O-V) model and is visualized in Figure~\ref{transactionflow}. Each phase and the different steps shown in Figure~\ref{transactionflow} are described below.
 
%\renewcommand{\labelenumi}{\Roman{enumi}}
%\renewcommand{\labelenumii}{\arabic{enumii}}
%\begin{enumerate}[wide, labelwidth=!, labelindent=0pt]
\paragraph*{Execution Phase}\mbox{}
% \begin{enumerate}[nosep]

\textbf{Step 1}: The client sends a transaction to all the endorsers. The transaction can include multiple reads and writes to one or more keys in the world state.

\textbf{Step 2}: The endorsers simulate the execution of the transaction on the world state and generate a read/write set that corresponds to the current world state of every key in the transaction. Then, the endorsers send a response back to the client that contains their own signature and the read/write set. This distributed execution of transactions on the endorsers helps to maintain trust without a centralized authority.

\textbf{Step 3}: The client collects the endorsing peers responses and sends them to the ordering service nodes. Optionally, the client may check the validity of the endorsing peers signatures and the consistency between the read/write set received from different peers. These are mandatorily checked later in the validation phase. Doing this check, the client can help detect transaction failures early in the transaction flow to reduce overhead.
%\end{enumerate}

\paragraph*{Ordering Phase}\mbox{}
%\subsection {Ordering Phase}
 
%\begin{enumerate}[nosep]
%\setcounter{enumi}{3}
\textbf{Step 4}:  The ordering service orders the transactions received from the client using a consensus protocol~\cite{lamport2001paxos, Ongaro:2014:SUC:2643634.2643666}. A transaction block is created based on three conditions: if a fixed duration of time has elapsed (\emph{block timeout}), if a fixed number of transactions have been received (\emph{block size}) or if the total size of transactions has reached a fixed limit (\emph{block max bytes}).

\textbf{Step 5}:  The block of transactions is then sent to all the peers.
%\end{enumerate}

\paragraph*{Validation Phase}\mbox{}

%\begin{enumerate}[nosep]
%\setcounter{enumi}{5}
\textbf{Step 6}:  Every peer, upon receiving a block of transactions from the ordering service, validates each transaction in the block independently. A peer checks if a sufficient number of valid endorsing peer signatures, based on the endorsement policy, have been collected (Validation System Chaincode (VSCC) validation). Then, the peer verifies if the version of every key in the read set of each transaction is equal to the version of the same key in the current world state (MVCC validation).

\textbf{Step 7}:  If VSCC and MVCC validation checks pass, the write sets of the transactions are applied to the world state. If any of the validation checks fail, the client is notified that the transaction aborted, and the world state does not change.

\textbf{Step 8}:  The validated block containing both aborted and committed transactions is appended to the ledger. The commit or abort status of every transaction is logged.
%\end{enumerate}

\section{Types of Transaction Failures}
\label{sec:failures}

{The endorsement phase and the other two phases (ordering and validation) happen in parallel. Hence, Fabric may execute transactions in the endorsement phase before previous transactions are ordered and committed. Thus, transactions are not always executed on the latest state of the world state in the endorsement phase. We have identified three types of transaction failures which are caused by this problem. Before formalizing them, we define basic concepts.}

\subsection{Basic Concepts}
The following definitions use the notations shown in Table~\ref{notations}.
 
%\begin{table}
%\small
%\centering
%\begin{tabular}[t]{ll} \hline
%P & set of endorsing peers \\
%T & set of all transactions \\
%B & set of blocks of transactions \\
%$\mathcal{K}^R$ & set of keys in a read set \\
%$\mathcal{V}^R$ & set of versions of keys in a read set \\
%$\mathcal{K}^W$ & set of keys in a write set \\
%$\mathfrak{V}^W$ & set of values of keys in a write set \\
%$\mathcal{K}^X$ & set of keys in the world state \\
%$\mathcal{V}^X$ & set of versions of keys in the world state \\
%$\mathfrak{V}^X$ & set of values of keys in the world state \\ \hline
%\end{tabular}
%\caption{Notations}
%\label{notations}
%\end{table}

\begin{table}
\footnotesize
\centering
\begin{tabular}[t]{ll} \hline
P: endorsing peers &  $\mathcal{V}^R$: versions of keys in a read set\\
T: transactions & $\mathfrak{V}^W$: values of keys in a write set \\
B: blocks of transactions &  $\mathcal{K}^X$: keys in the world state\\
$\mathcal{K}^R$: keys in a read set & $\mathcal{V}^X$: versions of keys in the world state \\
$\mathcal{K}^W$: keys in a write set  & $\mathfrak{V}^X$: values of keys in the world state\\  \hline
\end{tabular}
\caption{Notations for sets}
\label{notations}
\end{table}

%\footnote{All pairs are ordered to make the failure definitions more readable} 
\textbf{Definition 1: Read Set.} The read set $\mathit{RS_{T_iP_a}}$ of a transaction $T_i \in T$ generated by an endorsing peer $P_a \in P$ is represented by a set of $m \in \mathbb{N}_0$ ordered pairs of keys $\mathit{\mathcal{K}^R_k} \in \mathit{\mathcal{K}^R}$ and corresponding versions $\mathit{\mathcal{V}^R_k} \in \mathit{\mathcal{V}^R}$:
\begin{center}$\mathit{RS_{T_iP_a} = \{(\mathcal{K}^R_1, \mathcal{V}^R_1), \dots, (\mathcal{K}^R_m, \mathcal{V}^R_m)\}}$\end{center}

\textbf{Definition 2: Write Set.} The write set $\mathit{WS_{T_iP_a}}$ of a transaction $T_i \in T$ generated by an endorsing peer $P_a \in P$ is represented by a set of $n \in \mathbb{N}_0$ ordered pairs of keys $\mathit{\mathcal{K}^W_k} \in \mathit{\mathcal{K}^W}$ and corresponding values $\mathit{\mathfrak{V}^W_k} \in \mathit{\mathfrak{V}^W}$:
\begin{center}$\mathit{WS_{T_iP_a} = \{(\mathcal{K}^W_1, \mathfrak{V}^W_1), \dots, (\mathcal{K}^W_n, \mathfrak{V}^W_n)\}}$\end{center}

\textbf{Definition 3: World State.} The world state $\mathit{WX}$ is represented by an ordered set of $o \in \mathbb{N}_0$ keys $\mathit{\mathcal{K}^X_k} \in \mathit{\mathcal{K}^X}$, corresponding versions $\mathit{\mathcal{V}^X_k} \in \mathit{\mathcal{V}^X}$ and values $\mathit{\mathfrak{V}^X_k} \in \mathit{\mathfrak{V}^X}$:
\begin{center}$\mathit{WX = \{(\mathcal{K}^X_1, \mathcal{V}^X_1, \mathfrak{V}^X_1), \dots, (\mathcal{K}^X_o, \mathcal{V}^X_o, \mathfrak{V}^X_o)\}}$\end{center}

\textbf{Definition 4: Transaction Dependency.} A transaction $T_i$ is dependent on transaction $T_j$ if the read set $\mathit{RS_{T_iP_a}}$ of $T_i$  contains a key which is also present in the write set $\mathit{WS_{T_jP_a}}$ of $T_j$:
\begin{center}$\mathit{RS_{T_iP_a}  \cap WS_{T_jP_a} \neq \emptyset \, and \, i \neq j}$\end{center}

\subsection{Transaction Failures}

\subsubsection{Endorsement policy failures}: All transactions need to be endorsed by the endorsing peers in the execution phase (cf. Section~\ref{sec:background}). The endorsement of transactions can fail for multiple reasons, such as invalid endorser signatures or other technical reasons. Most of the possible causes for \emph{endorsement policy failures} are due to misconfigurations and unrelated to concurrency. In this study, we only consider endorsement policy failures caused by a read/write set mismatch, as described in the following.

The key-value store, which maintains the world state, is updated by each peer independently in the validation phase. Therefore, transient world state inconsistencies between the peers are possible. At the same time, the endorsing peers use the world state to generate read/write sets in the execution phase. Thus, the world state inconsistencies lead to a read/write set mismatch in the endorsement response causing an endorsement policy failure of the transaction. 
\begin{figure}[H]
\centering
\includegraphics[width=\columnwidth]{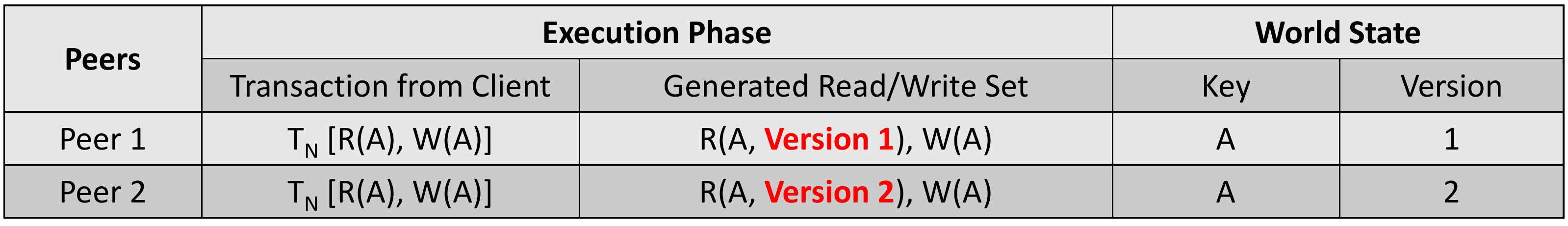}
\captionsetup{labelfont={bf}}
\caption{Example of an endorsement policy failure}
\label{EndorsementFailure}
\end{figure}
%{An example is illustrated in Figure~\ref{EndorsementFailure}. Here, Transaction 1 (${T_1}$) and Transaction 2 (${T_2}$) both update Key A and change its version to 1 and 2, respectively. In Peer 2, the validation of both ${T_1}$ and ${T_2}$ are complete while in Peer 1, only ${T_1}$ has been validated so far, i.e., \mbox{Peer 1} and Peer 2 have inconsistent world states.  At this point in time a new transaction ${T_N}$ which reads Key A is being executed on both Peer 1 and Peer 2. The result of this execution will give different results on both peers since the version of Key A is different. This results in an endorsement policy failure.}
{An example is illustrated in Figure~\ref{EndorsementFailure}. Here, the world states of Peer 1 and Peer 2 are inconsistent, i.e., the version of Key A is different on both peers. At this point in time a new transaction ${T_N}$ which reads Key A is being executed on both Peer 1 and \mbox{Peer 2}. This execution would give rise to different results on both peers since the version of Key A is different. This results in an endorsement policy failure.}  

Formally, an endorsement policy failure occurs when there exist two different endorsing peers $P_a, P_b \in P$ that both endorse the same transaction $T_i \in T$, such that there is the same read key $\mathit{\mathcal{K}^R_k \in \mathcal{K}^R}$ in the corresponding read sets $\mathit{RS_{T_iP_a}}$ and $\mathit{RS_{T_iP_b}}$, but the versions $\mathcal{V}^R_k$ of that key in $\mathit{RS_{T_iP_a}}$ and $\mathit{RS_{T_iP_b}}$ are different:
\begin{multline}
\begin{gathered}
\mathit{\exists \, T_i \in T \, \land \, P_a,P_b \in P \, \land \, \mathcal{K}^R_k \in \mathcal{K}^R \,\land \, \mathcal{V}^R_k \in \mathcal{V}^R} \\
\mathit{\land \, a \neq b \, such \, that \, \mathcal{K}^R_k \, of \, RS_{T_iP_a} == \mathcal{K}^R_k \, of \, RS_{T_iP_b} \, \land} \\
\mathit{\mathcal{V}^R_k \, of \, RS_{T_iP_a} \neq \mathcal{V}^R_k \, of \, RS_{T_iP_b}}
\end{gathered}
\end{multline}

\subsubsection{MVCC read conflicts}: \emph{MVCC read conflicts} are a well known problem in systems that have a multi-version view of each key~\cite{MURO1984207}. Every successful write on a key will increment the version and update the value. Every successful read will get the current version of the key in the world state along with its value. While a transaction moves from the execution phase to the validation phase, other transactions can get validated and committed, thereby updating the world state. Therefore, when a transaction reaches the validation stage, the world state may have changed and be different from the endorsement. 

Transactions that access the same key and thereby create a dependency are subject to \emph{MVCC read conflicts}. We further distinguish between \emph{inter-block MVCC read conflicts} and \emph{intra-block MVCC read conflicts}. Transaction failures caused due to a dependency among transactions in the same block are defined as intra-block MVCC read conflicts. If the cause is a dependency among transactions in different blocks, we call this inter-block MVCC read conflicts. 

Though the basic cause for both conflicts is the transaction dependency, other parameters such as block size influence these failures differently. For example, if two dependent transactions are submitted far apart in time and if they are included in different blocks, the first block may get committed before the second transaction is endorsed. Thus, both transactions could potentially succeed. {However, if in the same scenario, the block size was very large, they could get endorsed and then ordered into the same block. This would cause one of the transactions to fail validation.} A further difference is that intra-block MVCC read conflicts can potentially be resolved by transaction reordering~\cite{Sharma:2019:BLB:3299869.3319883} while inter-block MVCC read conflicts cannot be resolved in such a way. 
\begin{figure}[H]
\centering
\includegraphics[width=\columnwidth]{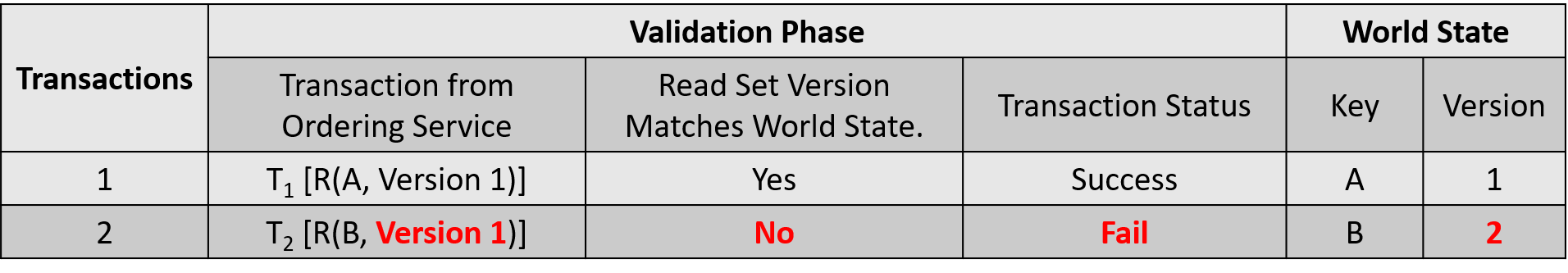}
\caption{Example of an MVCC read conflict}
\label{MVCC}
\end{figure}
{An example of MVCC read conflicts is illustrated in Figure~\ref{MVCC}. Here, Transaction 1 (${T_1}$) reads Key A whose version in the world state is the same as the version in the read set of the transaction. Hence, the read set contains the latest value of Key A. Whereas for Transaction 2 (${T_2}$) that reads Key B, there are different versions in the world state and the read set. This implies that ${T_2}$ is accessing an older version of the key and therefore fails.}

Formally, an MVCC read conflict occurs for a transaction when there exists a key $\mathit{\mathcal{K}^R_k \in \mathcal{K}^R}$ in the read set that matches a key $\mathit{\mathcal{K}^X_k \in \mathcal{K}^X}$ in the world state, but the version $\mathit{\mathcal{V}^R_k \in \mathcal{V}^R}$ of that key in the read set does not match the corresponding version $\mathit{\mathcal{V}^X_k \in \mathcal{V}^X}$ in the world state:
\begin{multline} \label{mvcc}
\begin{gathered}
\mathit{\exists \, \mathcal{K}^R_k \in \mathcal{K}^R \,\land \, \mathcal{K}^X_k \in \mathcal{K}^X \,\land \, \mathcal{V}^R_k \in \mathcal{V}^R\ \land \, \mathcal{V}^X_k \in \mathcal{V}^X} \\
\mathit{such \, that \, \mathcal{K}^R_k \, == \mathcal{K}^X_k \, \, \land \, \mathcal{V}^R_k  \neq \mathcal{V}^X_k}
\end{gathered}
\end{multline}
An MVCC read conflict is an intra-block conflict when \mbox{Equation \ref{mvcc}} holds and additionally, the cause of the MVCC read conflict is due to the dependency between two transactions $T_i$ and $T_j$ in the same block $\mathit{b_l \in B}$, i.e., the read set of $T_i$ contains a key $\mathit{\mathcal{K}^R_k}$ that is also present in the write set of $T_j$, but $T_j$ occurs before $T_i$ in the block:
\begin{multline}
\begin{gathered}
\mathit{\exists \, \mathcal{K}^R_k \in \mathcal{K}^R \,\land \, \mathcal{K}^W_k \in \mathcal{K}^W \,\land \, T_i,T_j \in T} \\
\mathit{\land \, T_i,T_j \in b_l \, \land \, b_l \in B  \land \, j < i}  \\
\mathit{such \, that \, \mathcal{K}^R_k \, of \, T_i \,== \mathcal{K}^W_k \, of \, T_j}
\end{gathered}
\end{multline}

Similarly, an MVCC read conflict is an inter-block conflict when Equation \ref{mvcc} holds and additionally, the MVCC read conflict is due to two transactions $T_i$ and $T_j$ in two different blocks $\mathit{b_l}$ and $\mathit{b_m}$, where the transaction $T_j$ that writes the key occurs in an earlier block than the transaction $T_i$ that reads the key:
\begin{multline}
\begin{gathered}
\mathit{\exists \, \mathcal{K}^R_k \in \mathcal{K}^R \,\land \, \mathcal{K}^W_k \in \mathcal{K}^W \,\land \,  T_i,T_j \in T \,\land \, T_i \in b_l} \\
\mathit{\,\land \, T_j \in b_m \,\land  \, b_l,b_m \in B  \land \, m < l} \\
\mathit{such \, that \, \mathcal{K}^R_k \, of \, T_i \,== \mathcal{K}^W_k \, of \, T_j}
\end{gathered}
\end{multline}

\subsubsection{Phantom read conflicts}: Fabric supports \emph{range queries} which read a set of keys from the key-value store. Given a start and end key, range  queries read all the keys from the world state within this interval.  In the validation phase, the entire range of keys is checked to ensure that no keys were inserted, deleted or updated within that range. If at least one key in the range has been inserted, deleted or updated, the transaction is aborted, and such failures are identified as \emph{phantom read conflicts}. 

Formally, a phantom read conflict occurs when Equation \ref{mvcc} holds or under one of the following conditions. First, when a key $\mathit{\mathcal{K}^R_k \in \mathcal{K}^R}$ in the read set is not present in the world state. Second, when there exists a key $\mathit{\mathcal{K}^X_k \in \mathcal{K}^X}$ in the world state  that is within the key interval $[i,j]$ of the range query, but is not present in the read set:
\begin{equation}
\begin{gathered}
\mathit{\exists \, \mathcal{K}^R_k \not\in \mathcal{K}^X \,\lor \, \mathcal{K}^X_k \not\in \mathcal{K}^R \, where \, i \,\leq \,k \,\leq\, j} \\
\end{gathered}
\end{equation}

Phantom read conflicts are essentially MVCC read conflicts since the cause for both is the dependency between transactions. However, Fabric identifies them separately since they are specific to range queries. The separate analysis of phantom read conflicts is justified, as we expect transactions with range queries to have more transaction dependencies than other transactions. For example, a range query that reads all the keys in the world state is dependent on every other write transaction.

\section{Experimental Methodology}
\label{sec:benchmark}

\subsection{HyperLedgerLab}
%~\cite{hyplab}
We developed \textsc{HyperLedgerLab}~\cite{fablab} as an automated solution for setting up the experimental environment. It initially uses OpenStack~\cite{openstack} APIs to commission the required infrastructure. Then, a Kubernetes~\cite{kuber} cluster is launched on this infrastructure. Next, the Fabric network is deployed on the cluster and the benchmark is started. The Fabric network can be deleted and recreated with different configurations to run the benchmark multiple times. The two benchmarks currently available for permissioned blockchains are Blockbench~\cite{Dinh:2017:BFA:3035918.3064033} and Hyperledger Caliper~\cite{caliper}. During the inception of this paper, Blockbench only supported Fabric \mbox{version 0.6}\footnote{Fabric 0.6 follows an Order-Execute model which is incompatible with the Execute-Order-Validate model of Fabric 1.4. As of April 2020 Blockbench supports Fabric 1.4.}. Therefore, we chose to integrate Caliper with \textsc{HyperLedgerLab}. It has a Fabric adapter that uses the \textsf{nodeJS-SDK} to communicate with the Fabric network via \textsf{gRPC}. Caliper monitors the performance of the blockchain and gathers metrics. We extended Caliper to collect all the different types of transaction failures necessary for our evaluation. We then developed four use-case based chaincodes and workloads which are explained in Section~\ref{chain} and a chaincode and workload generator which is explained in Section~\ref{synchain}.

\subsection{Cluster Setup}\label{cluster}

{We used a Kubernetes cluster consisting of one command line interface (CLI) node, three controller nodes, one load balancer, one network file system (NFS) node and multiple worker nodes. Every node runs on a Ubuntu Xenial (16.04) virtual machine. The CLI node and worker nodes have 16 vCPUs and 41 GB RAM each, the controller nodes have four vCPUs and 20 GB RAM each, and the load balancer and NFS node have two vCPUs and 10 GB RAM each. The workers host the pods for the different Fabric components such as peers and orderers. The controller nodes of the Kubernetes cluster are responsible for scheduling. The peers and orderers are deployed on the worker nodes in a round-robin fashion by the Kubernetes scheduler. }

{We use two different cluster setups for our experiments. The first setup (\emph{C1}) uses 3 worker nodes on which 4 peers and 3 orderers are launched and 5 client processes which are launched on the CLI node. The second setup (\emph{C2}) uses 32 worker nodes on which 32 peers and 3 orderers are launched. Also, in this setup 25 client processes are launched on the CLI node. Fabric 1.4 supports both the Solo and Kafka ordering service. We use Kafka for our experiments since Solo is not used in production.}

\begin{table}[!htb]
\caption{Chaincode functions and operations}
\label{cc}
\setlength{\tabcolsep}{0.5pt}
%\scriptsize
\footnotesize
\begin{minipage}{0.9\columnwidth}
\begin{tabular}[t]{|llll|} 
\hline
\multicolumn{4}{|c|}{\textbf{EHR}}\\ \hline
\rule{0pt}{\normalbaselineskip}\hspace{2mm}\textbf{Functions}&\hspace{2mm}\textbf{Operations}&\hspace{8mm}\textbf{Functions}&\hspace{2mm}\textbf{Operations}\hspace{2mm}\\
\hspace{2mm}initLedger&\hspace{2.5mm}2xW&\hspace{8mm}addEhr&\hspace{2.5mm}2xR, 2xW\\ 
\hspace{2mm}grantProfileAccess&\hspace{2.5mm}1xR, 1xW&\hspace{8mm}readProfile&\hspace{2.5mm}1xR\\ 
\hspace{2mm}revokeProfileAccess&\hspace{2.5mm}1xR, 1xW&\hspace{8mm}viewPartialProfile&\hspace{2.5mm}1xR\\
\hspace{2mm}revokeEhrAccess&\hspace{2.5mm}2xR, 2xW&\hspace{8mm}viewEHR	&\hspace{2.5mm}1xR\\ 
\hspace{2mm}grantEhrAccess&\hspace{2.5mm}2xR, 2xW&\hspace{8mm}queryEHR&\hspace{2.5mm}1xR\\ \hline
%addEhr&2xR, 2xW\\
%readProf&1xR\\ 
%vwParProf&1xR\\ 
%viewEHR	&1xR\\ 
%queryEHR&1xR \\ \hline
\end{tabular}
\end{minipage}
%\begin{minipage}{.49\columnwidth}
\begin{minipage}{.3\columnwidth}
\begin{tabular}[t]{|ll|} 
\hline
\multicolumn{2}{|c|}{\textbf{DV}}\\ \hline
\rule{0pt}{\normalbaselineskip}\hspace{0.5mm}\textbf{Functions}&\hspace{0.5mm}\textbf{Operations}\\ 
\hspace{0.5mm}initLedger&\hspace{0.5mm}3xW\\ 
%~&3 x W\\
\hspace{0.5mm}vote&\hspace{0.5mm}1xR, 2xRR,\\
~&\hspace{1mm}2xW\\
\hspace{0.5mm}closeElctn&\hspace{0.5mm}1xR, 1xW\\
\hspace{0.5mm}qryParties&\hspace{0.5mm}1xR, 1xRR\\
\hspace{0.5mm}seeResults&\hspace{0.5mm}1xR, 1xRR\\ \hline
%~&~ \\ \hline
\end{tabular}
\end{minipage}
%\begin{minipage}{.49\columnwidth}
\begin{minipage}{.3\columnwidth}
\begin{tabular}[t]{|ll|} 
\hline
\multicolumn{2}{|c|}{\textbf{SCM}}\\ \hline
\rule{0pt}{\normalbaselineskip}\hspace{0.5mm}\textbf{Functions}&\hspace{0.5mm}\textbf{Operations}\\
\hspace{0.5mm}initLedger&\hspace{0.5mm}2xW\\ 
\hspace{0.5mm}pushASN&\hspace{0.5mm}1xW\\ 
\hspace{0.5mm}Ship&\hspace{0.5mm}2xR, 2xW\\
\hspace{0.5mm}Unload&\hspace{0.5mm}2xR, 2xW\\
\hspace{0.5mm}queryASN&\hspace{0.5mm}1xRR\\
\hspace{0.5mm}queryStock&\hspace{0.5mm}1xRR*\\ \hline
\end{tabular}
\end{minipage}
%\begin{minipage}{.49\columnwidth}
\begin{minipage}{.3\columnwidth}
\begin{tabular}[t]{|ll|} 
\hline
\multicolumn{2}{|c|}{\textbf{DRM}}\\ \hline
\rule{0pt}{\normalbaselineskip}\hspace{0.5mm}\textbf{Functions}&\hspace{0.5mm}\textbf{Operations}\\
\hspace{0.5mm}initLedger&\hspace{0.5mm}2xW\\ 
\hspace{0.5mm}create&\hspace{0.5mm}1xR, 2xW\\
\hspace{0.5mm}play&\hspace{0.5mm}2xR, 1xW\\
\hspace{0.5mm}queryRghts&\hspace{0.5mm}2xR\\ 
\hspace{0.5mm}viewMetaData&\hspace{0.5mm}1xR\\ 
\hspace{0.5mm}calcRevenue&\hspace{0.5mm}1xRR*\\ \hline
\end{tabular}
\end{minipage}
%\multicolumn{2}{p{\textwidth}}\small{\textit{\hspace{4mm}*Fabric does not detect phantom reads for certain type of range reads (RR)}}
{*Fabric does not detect phantom reads for certain type of range reads (RR)}
\end{table}
%which are marked with an asterisk \textquotedblleft *\textquotedblright symbol in the table.
\subsection{Use-Case Based Chaincodes \& Workloads}\label{chain}

We developed four different chaincodes based on popular use cases from various disciplines to diversify our evaluation and make it realistic. The functions of each chaincode are shown in Table~\ref{cc} along with the number and type of read and write operations performed by each function. %There are two types of range read operations supported by Fabric. The results of the first type of range read operations are validated in the validation phase to detect phantom read conflicts. The result of the second type of range read operations are not validated and phantom read conflicts cannot be detected. In Table~\ref{cc}, we marked occurrences of the second type of range read operations additionally with an asterisk ``*'' symbol.Next, we discuss each chaincode and their workloads. 
All the read and write functions access keys randomly. The keys accessed by the range reads are described with each chaincode. For each chaincode, we initially populate the world state as described below. We intentionally used small numbers of keys in order to induce a high number of conflicts.

%A more detailed explanation of our chaincodes can be found in Ref.~\cite{hyperledgerlabchaincodes}. Comment Ruben: The reference is to unpublished work. Will that be submitted as supplementary material to the reviewers? If not, I would not refer to it.

\emph{Electronic Health Records (EHR)}: This chaincode manages medical health records provided by medical institutions or other service providers. Every patient owns two entities, its profile (personal information) and its electronic health records. We generate 100 profiles and 100 electronic health records to populate the world state. Access to either or both can be granted or revoked at any time. If access is granted, medical actors (doctors or researchers) may query or update the records. This chaincode only deals with the access credentials and logical connections. The actual data can be stored off chain. Similar designs are available in the literature~\cite{munoz2019clinicappchain, yang2018design, azaria2016medrec, mikula2018identity}.

\emph{Digital Voting (DV)}: A predefined set of 1000 voters and 12 competing parties participate in this digital voting scenario. Votes may be cast only during the election phase which ends with a close transaction. A voter is blocked from casting multiple votes. Querying the list of parties and counting out the votes are also included. The \texttt{qryParties} and \texttt{seeResults} functions query all 12 parties and the \texttt{vote} function queries all 1000 voters. This design is based on the work of Yavuz et al.~\cite{yavuz2018towards}.
%which is an Ethereum-based electronic voting application.

\emph{Supply Chain Management (SCM)}: This chaincode implements the standard operations of a general logistics network. Logistic service providers (LSP) and logistic units can be managed. For our workload, we generate five LSPs where four LSPs have 400 logistic units each and the fifth LSP has 800 logistic units. Global trade item numbers and serial shipping container codes are used to track the logistic units which can be single trade items or a group of items. An advanced shipping notice (ASN) can also be defined prior to a shipping. Upon successful shipping, the logistic unit is removed from the originating LSP and added to the destination LSP. Information regarding the units located at any LSP or information about a specific logistic unit can be retrieved. Also, any logistic unit can be unloaded to extract the embedded trade items. The \texttt{queryASN} function queries all the logistic units of a random LSP. This chaincode is based on the concepts of Perboli et al.~\cite{perboli2018blockchain}.

\emph{Digital Rights Management (DRM)}: This chaincode allows artists to share and manage their work on a blockchain. The metadata of 200 artworks is stored in the dot blockchain media format~\cite{dotbc} and 200 right holders can be identified with the industry standard IDs~\cite{csac}. Metadata and royalty management is handled on the blockchain. The current revenue of the right holders can also be calculated. There are similar blockchain applications in the market that handle music management and distribution ~\cite{verifi, ujo}.

\subsection{Synthetic Chaincodes \& Workloads}\label{synchain}
In order to run controlled experiments and microbenchmarks, we developed a chaincode and workload generator. The chaincode generator takes as input the total number of chaincode functions and for each function the number of read, insert, update, delete and range read actions. Users can also input the kind of database they wish to deploy; in case it is CouchDB, the user can also select to include rich queries in the chaincode functions. The final output is a syntactically correct chaincode with the user-specified chaincode functions. The workload generator takes as input the number of transactions, the transaction distribution (percentage of read, insert, delete, update and range read) and the key distribution (Zipfian skew). The output is a set of transactions based on the transaction and key distribution. In this paper, since we need to evaluate transaction failures caused by every transaction type, we generated a chaincode named \emph{genChain} which comprises equally distributed read, insert, update, delete and range read functions. {We initialized the world state with a large number of keys (100,000 keys) to run experiments with reduced transaction conflicts and generated read-heavy (RH), insert-heavy (IH), update-heavy (UH), delete-heavy (DH) and range-heavy (RaH) workloads. Each "x"-heavy (where x=read, insert, update, delete and range) workload contains 80\% of "x" transactions and a uniform distribution of the four other types of transactions. The range queries access a range of 2, 4 or 8 keys uniformly at random. We also generated a uniform workload of read and update transactions with 3 different key distributions (Zipfian skew: 0, 1, 2). }

\begin{table}[!htb]
\caption{Default values of control variables}
\label{controlvariablest}
\footnotesize
\centering
\renewcommand{\arraystretch}{1.2}
\begin{tabular}[t]{|l|l"l|l|} \hline
\textbf{Variable}&\textbf{Value}&\textbf{Variable}&\textbf{Value}\\ \hline
%\textbf{Variable}&~&\textbf{Variable}&~\\ \hline
Fabric Version&Fabric 1.4&Database Type&CouchDB\\ \hline 
Chaincode&EHR&Block Size&100\\ \hline
Policy&$P_0$&Tx arrival rate&100 tps\\ \hline 
%policy&~&arrival rate&~\\ \hline 
%Number of&2 (C1 cluster)&Number of peers&2 (C1 cluster)\\
%organizations&8 (C2 cluster)&per organization&4 (C2 cluster)\\ \hline 
No. of orgs&2 (C1); 8 (C2)&No. of peers / org&2 (C1); 4 (C2)\\ \hline
Zipfian Skew&1&Workload&Uniform\\ \hline 
\end{tabular}
\end{table}

\subsection{Control Variables \& Metrics}
We run our experiments using four different builds of Fabric. Fabric 1.4~\cite{fabcode} was the latest publicly available release version of Fabric during the inception of this paper. Fabric++~\cite{Sharma:2019:BLB:3299869.3319883}, Streamchain~\cite{istvan2018streamchain} and FabricSharp~\cite{10.1145/3318464.3389693} are extensions of Fabric that realize different optimization techniques. %Fabric++~\cite{pluscode} is an optimization of Fabric as described by Sharma et al.~\cite{Sharma:2019:BLB:3299869.3319883} where transaction reordering and early abort of transactions is implemented. Streamchain~\cite{streamcode} is another optimization of Fabric described by Istv{\'a}n et al.~\cite{istvan2018streamchain} where transactions are sent one by one instead of being batched into blocks.
The combined number of transactions sent per second from all clients is defined as the \emph{transaction arrival rate} of the system. \emph{Block size} is the number of transactions to be included in one block by the ordering service. We run experiments on the different \emph{chaincodes} (EHR, DV, SCM, DRM and genChain) as explained in Section~\ref{chain} and ~\ref{synchain}. The \emph{database type} can be set as CouchDB, which supports rich queries on values modeled as JSON data, or LevelDB, which is the default database for Fabric. 

\begin{figure*}[!htb]
\centering
\includegraphics[width=\textwidth]{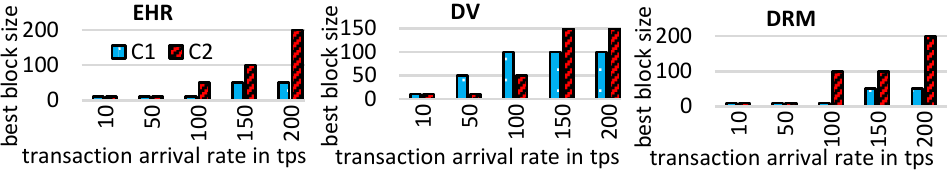}
\captionsetup{labelfont={bf}}
\caption{Best block size at different transaction arrival rates}
\label{bestblocksize}
\end{figure*}

\begin{figure*}[!htb]
\centering
\includegraphics[width=\textwidth]{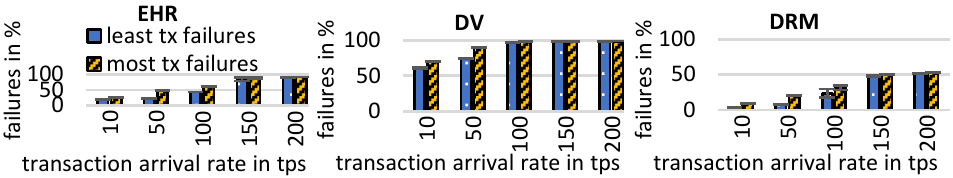}
\captionsetup{labelfont={bf}}
\caption{Minimum and maximum transaction failures}
\label{compfail}
\end{figure*}
\begin{figure*}[!htb]
\centering
\includegraphics[width=\textwidth]{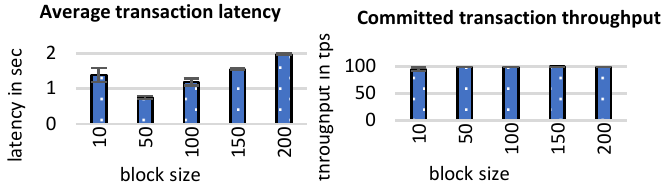}
\captionsetup{labelfont={bf}}
\caption{Latency and throughput at different block size}
\label{bslatthr}
\end{figure*}

The \emph{number of organizations}, \emph{number of peers per organization} and the \emph{endorsement policy} are other control variables. The different \emph{endorsement policies} we used are shown in Table~\ref{endpol}. We also vary the \emph{workload} to be read heavy, read-write heavy or uniform by changing the number of times each chaincode function is invoked. The keys accessed in our experiments follow a Zipfian distribution~\cite{powers1998applications} with varying \emph{Zipfian skew} values. A Zipfian skew of 0 implies that the keys are accessed uniformly. A positive Zipfian skew implies that the keys are accessed more from the higher range of the set of keys. We also do a network emulation with Pumba~\cite{pumba}, which is a chaos testing and network emulation tool for Docker containers, and induce a \emph{network delay} to evaluate different network conditions. Table~\ref{controlvariablest} shows the control variables and their default values. 

The performance metrics are collected by parsing the blockchain after each experiment. Therefore, the metrics collection process has no impact on the performance during the experiment. All failures are represented as percentages. {All failures including endorsement policy failures are detected only in the validation phase of Fabric. The clients do not resend any failed transactions and both failed and successful transactions are committed to the blockchain. The three extensions to Fabric have some exceptions to this default flow, which are mentioned in the corresponding sections. \emph{Average total transaction latency} is the average of the time taken for each complete transaction flow (all three phases of the E-O-V model) of both failed and successful transactions. \emph{Committed transaction throughput} is the number of transactions committed to the blockchain divided by the total time taken. }

\section{Experimental Results}
\label{sec:evaluation}
We conducted a total of {970 experiments}. Each experiment was repeated at least 3 times and the average result is presented here. Transactions were sent from all clients for a duration of three minutes for every experiment. With both the cluster configurations, our testbed yields a throughput of 200 tps (transactions per second). The default values of the control variables are defined in Table~\ref{controlvariablest} and changes to any of these values are explicitly mentioned with the results.

\subsection{Results for Fabric 1.4}

\subsubsection{Block size and transaction arrival rate}: 

\textbf{(a) Transaction Failures}\label{txfail}: We define the \emph{best block size} as the block size at which there is the least percentage of failed transactions and the \emph{worst block size} as the block size at which the percentage of failed transactions is maximum. Figure~\ref{bestblocksize} shows the \emph{best block size} at different transaction arrival rates for different chaincodes {and cluster setups}. Figure~\ref{compfail} shows the maximum and minimum percentage of failed transactions corresponding to the \emph{best block size} and \emph{worst block size} {on the C2 cluster. Figure~\ref{bslatthr} shows the effect of block size on latency and throughput at an arrival rate of 100 tps for the EHR on the C2 cluster.}

\textbf{Observations}: For all the chaincodes, we see an approximately linear relation between increasing {transaction arrival rate} and the {best block size}. There is up to 60\% decrease in {failed transactions} between the \emph{worst block size} and the \emph{{best block size}}. For example, the DRM chaincode at 50 tps yielded 21.14\% failures with the \emph{worst block size} while we observed only 8.07\% failures with the \emph{{best block size}}.

At low transaction arrival rates, a low block size is preferable, so that the ordering service does not have to wait for a long time until enough transactions have arrived for a block to be created. However, as the transaction arrival rate increases, this delay becomes less significant. Hence, at higher transaction arrival rates, larger blocks can be built. Building larger blocks has the advantage that there is less overhead involved in the ordering service and in the validating peers. This reduces the chance of temporary overload and queuing, leading to lower transaction latency and, hence, to less MVCC read conflicts. {We can also observe that due to more resources the C2 cluster setup supports higher block sizes at high transaction rates than the C1 cluster setup. To further understand the effect of block size, we also analyzed the latency and throughput. The \emph{best block size} for EHR at 100 tps is 50, the latency for EHR is lowest at a block size of 50 and the throughput is not significantly affected by block size (Figure~\ref{bslatthr}). Similarly, the \emph{best block size} at 150 tps is 100 and the highest throughput and least latency are also for block size 100 (not shown in the figure). So, the block size where other performance metrics such as latency and throughput have better values is also where the failures are least. We made similar observations with the other chaincodes at different transaction arrival rates.}
%Query processing systems like OLTPShare\cite{10.14778/3229863.3229866} and SharedDB\cite{10.14778/2168651.2168654} that batch queries have made similar observations about the relationship between batch sizes and performance.

The number of MVCC read conflicts also depends on the number of keys to validate which varies depending on the chaincode functions. Hence, the best setting of the block size with increasing transaction arrival rate is different for different chaincodes. Three of the five chaincode functions in DV have range queries which cause a higher failure rate when compared to the other chaincodes. We can still observe the influence of block size on transaction failures, but the effect is less significant than for the other chaincodes. 

%A block size less than or equal to the transaction arrival rate will help faster creation of blocks since the ordering service does not have to wait for sufficient transactions to create a block. If the block size is significantly less than the transaction arrival rate then too many blocks are created and congests the system. This increases the number of inter-block MVCC conflicts. On the other hand, if the block size is large then the number of intra-block MVCC read conflicts increases. Therefore we observe that the best block size does increase with increasing transaction arrival rate but it increases at a lower rate than the increase in transaction arrival rate. 

%Smaller blocks can be validated and committed faster since there are fewer transactions within a block. This reduces the overall latency of the system and consequently reduces the number of transaction failures. However, at higher transaction arrival rates, blocks are created faster than they are validated and committed. This causes blocks to queue up and increases the latency. That is why larger block sizes are preferable at higher transaction arrival rates. 

\textbf{Implications}: The percentage of {failed transactions} that occur when the {transaction arrival rate} changes depends on the block size in most cases. This dependency changes with different {chaincodes} {and cluster setups}. Determining this dependency for each {chaincode} on a specific Fabric network and adapting the block size when the {arrival rate} changes is an efficient and simple approach to reduce transaction failures. {Further, the measured latency and throughput comprise of both successful and failed transactions. So having low latency or high throughput is irrelevant if the transaction failure rate is very high. Therefore, transaction failures should be analyzed along with latency and throughput to ensure good performance.} %But changing the block size does not have a significant effect when a large number of range queries are present. 

\begin{figure*}[!htb]
\minipage{0.49\textwidth}
\includegraphics[width=\textwidth]{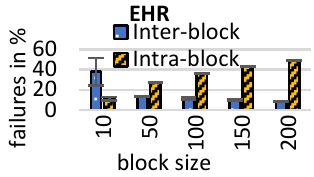}
\captionsetup{justification=centering} 
\captionsetup{labelfont={bf}}
\caption{Effect of block size}
%\caption{Effect of the \mbox{number} of organizations}
\label{interintra}
\endminipage\hfill
\minipage{0.49\textwidth}
\includegraphics[width=\textwidth]{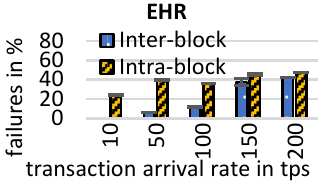}
\captionsetup{justification=centering} 
\captionsetup{labelfont={bf}}
\caption{Effect of load}
\label{interintra2}
\endminipage\hfill
\end{figure*}

\textbf{(b) MVCC read conflicts}: We evaluate the number of {inter-block MVCC read conflicts} and {intra-block MVCC read conflicts} with changing block size and transaction arrival rate in Figures~\ref{interintra} and ~\ref{interintra2}, respectively, for the EHR chaincode {on the C2 cluster}. 

\textbf{Observations}: The number of {intra-block MVCC read conflicts} increases when the block size increases because when more transactions are included in a block, there is a higher chance of dependencies between transactions which lead to conflicts. Conversely, with increasing block size the {inter-block MVCC read conflicts} decrease because the conflicts are more likely to have already occurred within the block than across blocks. Also, both failures increase with increasing transaction arrival rate. {Further, the \emph{best block size} for EHR at 100 tps is 50, and the sum of both inter and intra-block MVCC read conflicts are low at this block size (Figure~\ref{interintra}). But if we consider them individually, inter-block conflicts are least at block size 200 while intra-block conflicts is least at block size 10.}
%For SCM, these trends are less uniform because the percentage of MVCC read conflicts is lower.

\textbf{Implications}: {Changing the {block size} with respect to any one type of failure is not useful since different types of failures have a different relation to block size. A trade-off between the least inter-block and intra-block MVCC read conflicts is necessary to ensure that the total transaction failures are least.}

%\textbf{(b) MVCC read conflicts}: Figure 8 shows the block size at which the MVCC read conflicts (Inter and intra) are least at different transaction arrival rates. Figure 9 shows the highest and lowest percentage of MVCC  read conflicts (Inter and intra) at different transaction arrival rates. Figure 11 shows the effect of tps and blocksize on MVCC read conflicts.

%\textbf{Observations}: We observe from figure 8 and 9 that blocksize has a significant impact on the MVCC read conflicts. Intra mvcc conflicts are least at the lowest block size (10) while inter mvcc conflicts are lower at larger block sizes. This is because when the block size is larger (more transactions within each block) the probability of transaction dependencies within the block is higher. We observe from figure 11 that MVCC conflicts (inter and intra) increase with increase in transaction arrival rate. We again observe that the number of {intra-block MVCC read conflicts} increases when the block size increases. This is because when more transactions are included in a block, there is a higher chance of dependencies between transactions which leads to conflicts. Conversely, with increasing block size the {inter-block MVCC read conflicts} decrease because the conflicts are more likely to have already occurred within the block than across blocks. 

%\textbf{Implications}: Changing the {block size} with respect to any one type of failure is not useful since different types of failures have a different relation to block size. It is necessary to observe all types of transaction failures and their interrelation.

\begin{figure*}[!htb]
\minipage{0.49\textwidth}
\includegraphics[width=\textwidth]{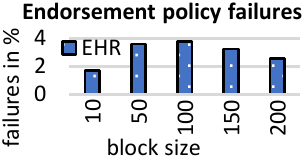}
\captionsetup{justification=centering} 
\captionsetup{labelfont={bf}}
\caption{Endorsement policy failures}
%\caption{Effect of the \mbox{number} of organizations}
\label{end}
\endminipage\hfill
\minipage{0.49\textwidth}
\includegraphics[width=\textwidth]{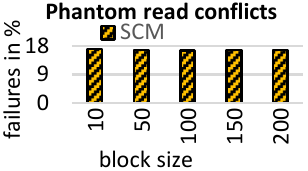}
\captionsetup{justification=centering} 
\captionsetup{labelfont={bf}}
\caption{Phantom read conflicts}
\label{phan}
\endminipage\hfill
\end{figure*}

\textbf{(c) Endorsement policy failures and phantom reads}: {Figure~\ref{end} and Figure~\ref{phan} show the effect of block size on endorsement policy failures for the EHR chaincode and phantom reads for the SCM chaincode, respectively, on the C2 cluster.}

\textbf{Observations}: {Since endorsement policy failures are caused by inconsistent world states, block size does not have a significant impact. A single range query transaction can have a dependency with multiple other transactions within and across blocks that write to at least one key contained in the range.}

\textbf{Implications}: {Though adapting the block size can help in reducing transaction failures, some types of failures like endorsement failures and phantom reads are not affected.}

\begin{figure*}[!htb]
\centering
\includegraphics[width=\textwidth]{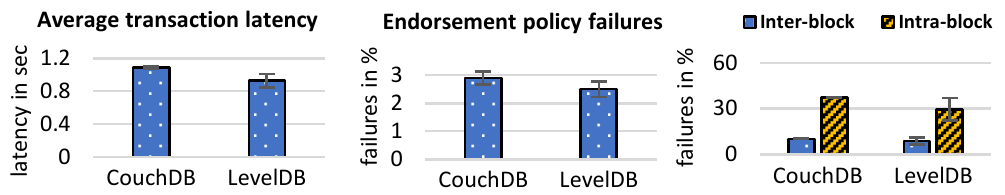}
\captionsetup{labelfont={bf}}
\caption{Effect of database type on latency and failures}
\label{db}
\end{figure*}

\begin{table*}[!htb]
\centering
\caption{Effect of database type}
\label{dbgen}
%\footnotesize
\begin{tabular}[t]{|c|c|c"c|c|c"c|c|c|} \hline
\multicolumn{3}{|c"}{\textbf{Average transaction latency (s)}}&\multicolumn{3}{c"}{\textbf{Transaction failures (\%)}}&\multicolumn{3}{c|}{\textbf{Function call latency (ms)}}\\ \hline
\textbf{Workload}&\textbf{CouchDB}&\textbf{LevelDB}&\textbf{Workload}&\textbf{CouchDB}&\textbf{LevelDB}&\textbf{Function}&\textbf{CouchDB}&\textbf{LevelDB}\\ \hline
ReadHeavy&18.04&3.22&ReadHeavy&5.65&1.38&GetState&8.3&0.6\\ \hline
InsertHeavy&18.34&7.93&InsertHeavy&2.17&1.36&PutState&0.8&0.5\\ \hline
UpdateHeavy&20.82&9.86&UpdateHeavy&31.31&23.03&GetRange&88&1.4\\ \hline
RangeHeavy&101.63&4.14&RangeHeavy&34.18&5.19&DeleteState&1.2&0.6\\ \hline
DeleteHeavy&18.48&1.22&DeleteHeavy&1.11&0.18&&&\\ \hline
\end{tabular}
\end{table*}

%\begin{figure}[H]
%\minipage{0.66\columnwidth}
%\includegraphics[width=\columnwidth]{figures/figure12.pdf}
%\captionsetup{justification=centering} 
%\captionsetup{labelfont={bf}}
%\caption{Effect of database type at different workloads}
%\caption{Effect of the \mbox{number} of organizations}
%\label{dbgen}
%\endminipage\hfill
%\minipage{0.34\columnwidth}
%\includegraphics[width=\columnwidth]{figures/figure13.pdf}
%%\captionsetup{justification=centering} 
%\captionsetup{labelfont={bf}}
%\caption{Function call latency}
%\label{dbfunc}
%\endminipage\hfill
%\end{figure}

\subsubsection{Database type}: The effect of using CouchDB (CDB) or LevelDB (LDB) with uniform workload and the EHR chaincode is shown in Figure~\ref{db} and the results with different workloads with the \emph{genChain} chaincode are shown in Table~\ref{dbgen}. It also shows the latency of the different chaincode function calls on both databases. 

\textbf{Observations}: LevelDB performs better across different chaincodes and workloads. {To better understand the overhead of CouchDB, we further analyze the latency of each different function call in the chaincode (Table~\ref{dbgen}). Latency is lower when using LevelDB because it is embedded with the peer process whereas CouchDB is an external database invoked via REST APIs~\cite{couchdbfabric}.} The percentages of {endorsement policy failures} and {MVCC read conflicts} are also lower with LevelDB. A lower latency implies that transactions are committed faster. This leads to fewer conflicts between transactions, i.e., {MVCC read conflicts} are reduced. Further, the world state can be updated faster which leads to a slight reduction of endorsement policy failures. 
{Also, the latency and correspondingly the number of failures for a range-heavy workload is significantly higher for CouchDB (Table~\ref{dbgen}). This is because the entire range of keys are read from the database during the endorsement phase and validation phase to ensure that no key has changed between the phases (phantom read detection). For an external database such as CouchDB, this induces a significant overhead. }
%\cite{8526892} measures the endorsement, ordering and validation latency of Hyperledger Fabric indivudually 

\textbf{Implications}: CouchDB supports rich queries such as sort and filter which are useful for many use cases, while LevelDB only supports simple get and set queries. Though CouchDB has richer functionality, our results show that it affects the performance of Fabric. So, if a chaincode can be designed without rich queries, it is always better to use LevelDB to reduce transaction failures. {Also, rich queries supported by CouchDB can provide similar functionality as a range query, but Fabric does not re-execute a rich query in the final validation phase and therefore, provides no guarantees on the validity of the query result (no phantom read detection)~\cite{couchquery}. So the user needs to make a trade-off between performance and query result validity.}

\begin{figure*}[!htb]
\centering
\includegraphics[width=\textwidth]{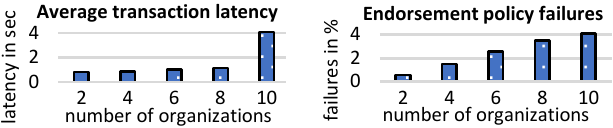}
\captionsetup{labelfont={bf}}
\caption{Effect of the number of organizations}
\label{org}
\end{figure*}

\subsubsection{Number of organizations}\label{orgsec}: Figure~\ref{org} shows the effect of the {number of organizations} on latency and transaction failures. Experiments were conducted with different numbers of organizations {(2, 4, 6, 8 and 10) on the C2 cluster. There are four peers per organization} and therefore increasing the number of organizations increases the number of peers involved in the system.

\textbf{Observations}: The transaction latency and the endorsement policy failures increase with the {number of organizations}. When there are more peers, the number of replicas of the world state increases, which in turn increases the possibility of inconsistent world states between the peers which causes the observed increase in the number of endorsement policy failures. %Since the transactions need to be validated and committed on more peers, the world state is updated slower. This leads to more {MVCC read conflicts}.

\textbf{Implications}: The number of organizations and peers that form a Fabric network should be restricted as much as possible to reduce endorsement policy failures. For example, geographically close or functionally similar branches of a company could be considered as a single organization. The trade-off between performance and inclusion of more participants should be considered when building a Fabric network.

\begin{table*}[htbp]
%\footnotesize
\centering
\caption{Endorsement Policies}
\label{endpol}
\begin{tabular}[t]{ll} \hline
$P_0$: \textquotedbl N-of"{}: [{ "signed-by": 0 }, ..., { "signed-by": N-1 }] \\
$P_1$: "2-of": [{ "signed-by": 0}, { "1-of": [{ "signed-by": 1 }, ..., { "signed-by": N-1 }]} ] \\
$P_2$: "2-of": [{ "1-of": [{ "signed-by": 0 }, ..., { "signed-by": N/2 }]}, \\ \hspace{20mm}{ "1-of": [{ "signed-by": N/2+1 }, ..., { "signed-by": N-1 }]}] \\
$P_3$: "(N/2+1)-of": [{ "signed-by": 0 }, ..., { "signed-by": N-1 }] \\
N: number of organizations, "n-of": n signatures required \\
An "n-of" clause nested inside another "n-of" is called a sub-policy. \\ \hline
\end{tabular}
\end{table*}

\begin{figure*}[!htb]
\centering
\includegraphics[width=\textwidth]{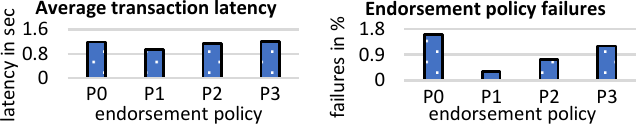}
%\captionsetup{labelfont={bf}}
\caption{Effect of endorsement policies}
\label{pol}
\end{figure*}

\subsubsection{Endorsement Policy}\label{polsec}: Figure~\ref{pol} shows the impact of different endorsement policies (cf. Table~\ref{endpol}). 

%The percentage of {endorsement policy failures} is influenced by the {endorsement policy}. 
\textbf{Observations}: The number of {endorsement policy failures} is maximal when the most endorsement signatures are required. Endorsement policy $P_0$ requires $N$ signatures and $P_3$ requires a quorum $(N/2+1)$ of signatures. When more endorsements are required, the world state needs to be consistent on a greater number of peers. Therefore, there is a higher chance for endorsement policy failures.

Even when an equal number of signatures is required, an increase in the number of sub-policies induces an increasing number of failures. For instance, $P_1$ and $P_2$ both require two signatures. $P_1$ requires one signature from organization $\mathit{Org}_0$ and one signature from any of the other organizations, whereas $P_2$ requires one signature from the first half of the organizations and one signature from the second half of the organizations. Therefore, $P_1$ includes one sub-policy while $P_2$ includes two sub-policies. Consequently, the number of endorsement policy failures in $P_2$ is higher. The endorsement policy is parsed during the VSCC validation and compared with the endorsement signatures of a transaction. Each sub-policy is a separate search space, so that the time taken for validation increases with an increasing number of sub-policies, which in turn also increases the chance of endorsement policy failures. At the same time, the average transaction latency increases.

%. Since both the validation phase and the execution phase occur on the endorsers the chance for inconsistent world state is higher when the latency of either phase increases.  
\textbf{Implications}: Similar to restricting the number of organizations and peers participating in the network, enterprises should also restrict the number of participants in the endorsement policy. For example, if one organization has a higher decision-making power or is more trustworthy than another, then one can reconsider if really both organizations are required to endorse the transactions. Also, one can consider simplifying endorsement policies, such that the number of sub-policies is reduced. For example, the policy: 
\begin{center}
"4-of": ["2-of": [$\mathit{Org_0}$, $\mathit{Org_1}$], "2-of": [$\mathit{Org_2}$, $\mathit{Org_3}$]]
\end{center}
can also be written as:
\begin{center}
"4-of": [$\mathit{Org_0}$, $\mathit{Org_1}$, $\mathit{Org_2}$, $\mathit{Org_3}$]
\end{center}
In either formulation, all four organizations have to endorse the transactions.

\begin{figure*}[!htb]
\minipage{0.49\textwidth}
\includegraphics[width=0.9\textwidth]{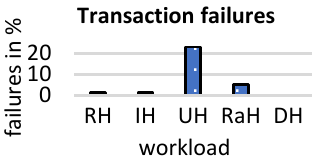}
\captionsetup{labelfont={bf}}
\caption{Effect of workload}
\label{workload}
\endminipage\hfill
\minipage{0.49\textwidth}
\includegraphics[width=0.9\textwidth]{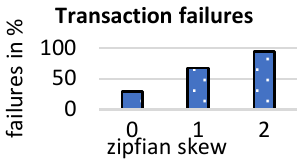}
\captionsetup{labelfont={bf}}
\caption{Effect of key skew}
\label{skew}
\endminipage\hfill
\end{figure*}

\subsubsection{Workload}: Figure~\ref{workload} shows the effect of different workloads on failures with \emph{genChain} chaincode {on the C2 cluster}.

%\textbf{Observations}: Read-heavy workloads induce the least failures and read-write heavy workloads induce the most failures. This is because only write transactions cause dependencies with reads that lead to conflicts. This result is independent of the block size.

\textbf{Observations}: 
Insert-heavy and delete-heavy workloads insert and delete unique keys, thus avoiding transaction conflicts. Hence, these workloads have the least failures. Read-heavy and range-heavy workloads induce lower failures compared to update-heavy workloads because only write transactions cause dependencies with reads that lead to conflicts. This result is independent of block size.

%\begin{figure}[tbp]
%\centering
%\includegraphics[width=\columnwidth]{figures/skewnewsingle}
%\caption{Effect of key access skew}
%\label{skew}
%end{figure}

\textbf{Implications}: If the use case permits, one should aim to batch read-only transactions together to ensure they all succeed. Our results also back the recommendation by Fabric~\cite{query} to not submit read-only transactions for ordering and validation since the necessary result is already delivered after the execution phase itself. It is only necessary to submit read-only transactions if one needs a record on the blockchain for auditing purposes.

\subsubsection{Zipfian skew}: Figure~\ref{skew} shows the effect of the {Zipfian skew} for key access with \emph{genChain} chaincode and a uniform workload.

\textbf{Observations}: When the key access is more skewed, the percentage of failures increases. The number of conflicts will increase if more transactions access the same (set of) key(s). 

\textbf{Implications}: Chaincodes and the database structure can be designed such that key access is less skewed. For example, in the EHR chaincode, the \texttt{addEHR} function uses a \texttt{PatientID} as the key to add any new medical record for a patient. One could replace \texttt{patientID} by two new keys \texttt{PatientID\_XrayID} and \texttt{PatientID\_MRIID} such that a transaction updating a patient's Xray and another transaction updating the same patient's MRI will not conflict with each other. Overall, modelling the data representation is an important aspect that should be carried out carefully.

\begin{figure*}[!htb]
\centering
\includegraphics[width=\textwidth]{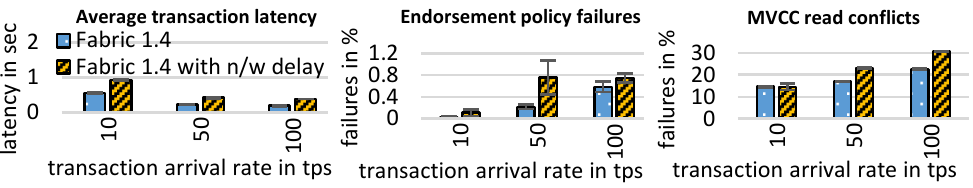}
\caption{Fabric 1.4 with and without network delay}
\label{network}
\end{figure*}
%\begin{figure}[H]
%\centering
%\includegraphics{figures/figure19(new).pdf}
%\caption{Fabric 1.4 with and without network delay}
%\label{network2}
%\end{figure}

\begin{figure*}[!htb]
\centering
\includegraphics[width=0.95\textwidth]{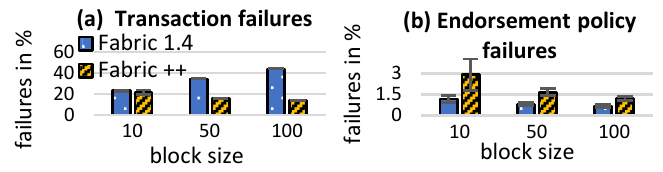}
\caption{Effect of block size}
\label{plusbs}
\end{figure*}

\begin{figure*}[!htb]
\centering
\includegraphics[width=0.95\textwidth]{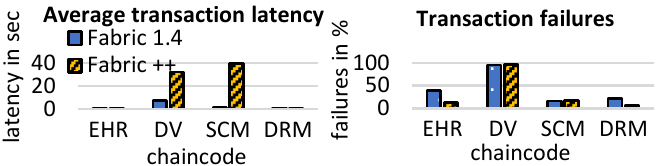}
\captionsetup{labelfont={bf}}
\caption{Effect of different chaincodes on Fabric++}
\label{pluschaincode}
\end{figure*}

\begin{figure*}[!htb]
\centering
\includegraphics[width=0.95\textwidth]{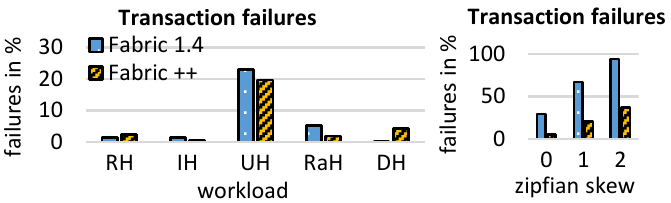}
\captionsetup{labelfont={bf}}
\caption{Effect of different workloads on Fabric++}
\label{plusgen}
\end{figure*}

\subsubsection{Network delay}: Figure~\ref{network} shows the effect of an induced additional network delay of 100$\pm$10 ms for one organization to emulate the scenario of a geographically distributed organization. 

\textbf{Observations}: The additional network delay causes an increase in the  transaction latency and consequently increases the number of failures. The {endorsement policy failures} are affected because the network delay increases the chance of inconsistent world states between peers. MVCC read conflicts are affected because the time between endorsement and validation of a transaction is increased.

\textbf{Implications}: Endorsement policy failures are highly affected by network delays. This implies that if several geographically far apart organizations are part of the endorsement policy, the number of endorsement policy failures will increase. Hence, network delays must be considered in the design of the Fabric network and of the endorsement policies. Finally, MVCC read conflicts that are caused by large network delays are inherent to the optimistic concurrency control in Fabric and cannot be avoided.

\subsection{Results for Fabric++}
Sharma et al.~\cite{Sharma:2019:BLB:3299869.3319883} designed an optimized extension of Fabric which can effectively resolve {intra-block MVCC read conflicts}. In the ordering phase of Fabric++, a conflict graph is generated for the transactions in a block and all cycles in the graph are identified. Cycles are removed from the graph by aborting transactions in the ordering phase itself. The resulting acyclic graph of transactions is serialized and sent to the validation phase.

\subsubsection{Block Size}: Figure~\ref{plusbs} (a) compares the percentage of {transaction failures} for Fabric 1.4 and Fabric++ at different {block sizes}.

\textbf{Observations}: At a fixed {transaction arrival rate} with Fabric++, the {transaction failures} decrease with increase in block size whereas the trend is reverse for Fabric 1.4. A larger block size gives more reordering possibilities for Fabric++ which leads to fewer {failures}.

\textbf{Implications}: To efficiently utilize the transaction reordering implemented by Fabric++, use a larger {block size}.

\subsubsection{Endorsement policy failures}: Figure~\ref{plusbs} (b) compares the percentage of {endorsement policy failures} in Fabric 1.4 and Fabric++.

\textbf{Observations}: The {endorsement policy failures} are higher for Fabric++ because there are less transaction failures due to MVCC read conflicts. Hence, the rate of updates of the world state is higher, because only successful transactions are committed to the world state. As the rate of updates is higher, there is also a higher chance of inconsistencies between the world state replicas on the different peers, which leads to more endorsement policy failures.

\textbf{Implications}: Reordering cannot resolve {endorsement policy failures}. This problem has to be investigated and treated separately, e.g., by reconsidering the endorsement policy or the design of the Fabric network.

%\subsubsection{Database type}: Figure~\ref{plusdb} compares the percentage of transaction failures in Fabric++ when using CouchDB and LevelDB.

%\textbf{Observations}: Similar to Fabric 1.4, Fabric++ achieves better results with LevelDB than with CouchDB. 

%\textbf{Implications}: In terms of performance, LevelDB is a better choice for Fabric++. Similar to Fabric, the lower latency of update operations decreases the number of transaction failures.

\subsubsection{Chaincodes {and Workloads}}\label{fabric++sec}: {Figures~\ref{pluschaincode} and ~\ref{plusgen} compare the latency and percentage of transaction failures in Fabric 1.4 and Fabric++ with different chaincodes, workloads and key distribution.}

\textbf{Observations}: The total failures do not significantly decrease with Fabric++ when evaluated with the DV and SCM chaincodes. These two chaincodes include range queries that cause {phantom read conflicts}. {Each range query involves a large range of keys (800 to 1000 keys)}, so that transactions will have dependencies on multiple other transactions. The latency is significantly higher for Fabric++ with these chaincodes. Fabric++ creates conflict graphs and then makes them acyclic to resolve transaction conflicts. To this end, they approximate a solution to the Minimum Feedback Vertex Set (MFVS) problem (which is an NP-hard problem). Because of the high number of dependencies induced by range queries, generating the conflict graph and reordering the range queries become very time consuming in Fabric++. {With the \emph{genChain} chaincode, Fabric++ reduces transaction failures for most of the workloads. Since the range queries in the range-heavy workload have a smaller range (2, 4 and  8), we observe a reduction in failures even in the presence of range queries. Further, Fabric++ does not have a positive effect on read-heavy and delete-heavy workloads because although the reordering possibilities are few for these workloads, the reordering process is still being executed and this increases the latency.}

\textbf{Implications}: Fabric++~\cite{Sharma:2019:BLB:3299869.3319883} did not evaluate the system in the presence of large range queries. Therefore, our results provide new insights on the impact of range queries on Fabric++. {If the use case permits, one should consider designing chaincodes with smaller range queries when using Fabric++.} Fabric++ could be optimized in the future to handle range queries more efficiently, e.g., by using a different algorithm to tackle the reordering problem. {Also, the reordering potential of a workload needs to be analyzed to efficiently use Fabric++. }
%If larger range queries need to be used, one should avoid using Fabric++, since the transaction latency will be high. 
\subsection{Results for Streamchain}
Streamchain~\cite{istvan2018streamchain} is an extension of Fabric that focuses on reducing the latency by sending transactions one-by-one instead of creating a block. In the validation phase, parallel validation of signatures and pipelining are implemented. The current prototype requires that the ledger and the world state are stored on a RAM disk both in the ordering service and the peers.

\begin{figure*}[!htb]
\centering
\includegraphics[width=0.95\textwidth]{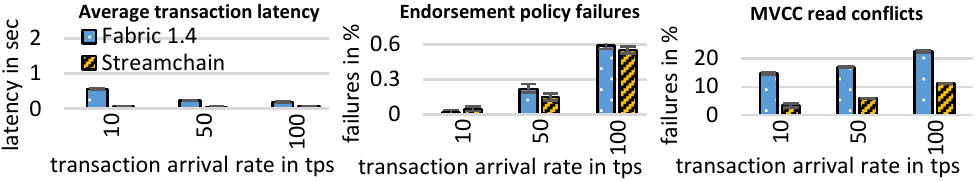}
\caption{Comparison of Streamchain and Fabric 1.4}
\label{streamchain}
\end{figure*}

\begin{figure*}[!htb]
\centering
\includegraphics[width=0.95\textwidth]{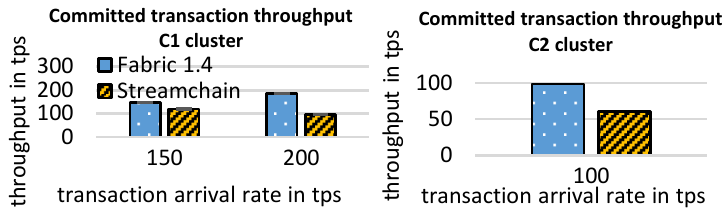}
\captionsetup{labelfont={bf}}
\caption{Latency and committed transaction throughput}
\label{stream150200}
\end{figure*}

\subsubsection{Latency and transaction failures}: Figure~\ref{streamchain} compares the performance of Streamchain and Fabric 1.4 at transaction rates of 10, 50 and 100 tps. Figure~\ref{stream150200} shows the committed transaction throughput for both at higher  arrival rates of 150 and 200 tps {on the C1 cluster and 100 tps on the C2 cluster}. Fabric 1.4 is set with a block size of 10 (we observed similar results with block sizes 50 and 100).

\textbf{Observations}: The latency and transaction failures are lower for Streamchain up to a transaction arrival rate of 100 tps {on the C1 cluster and up to 50 tps on the C2 cluster}. Since the transactions are streamed one-by-one and stored on a RAM disk, the world state is updated quickly, thus reducing the MVCC read conflicts. Since the latency is lower, endorsement policy failures also reduce slightly. Beyond a transaction arrival rate of 150 tps {on the C1 cluster}, Streamchain does not provide enough throughput to handle the load. {Further, on the larger C2 cluster the overhead is prominent even at an arrival rate of 100 tps. Streaming the transactions one-by-one will increase the communication overhead between the orderer and the multiple peers. At higher transaction rates and with larger number of peers (C2 cluster), this results in queuing of transactions.}
%Hence, the latency increases significantly.

\textbf{Implications}: Streaming the transactions one-by-one helps to update the world state faster at low transaction arrival rates. But Streamchain needs to be further optimized to handle high transaction arrival rates {and scaling}. 

\begin{figure*}[!htb]
\centering
\includegraphics[width=0.9\textwidth]{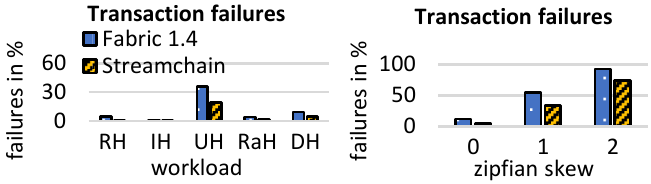}
\captionsetup{labelfont={bf}}
\caption{Effect of different workloads on Streamchain}
\label{streamgen}
\end{figure*}

\subsubsection{{Workloads}}: {Figure~\ref{streamgen} compares the performance of Streamchain and Fabric 1.4 with different workloads and key distribution at 50 tps on the C2 cluster.}

\textbf{Observations}: {Streamchain reduces the transaction failures regardless of the type of workload or key distribution. This is because the optimization used by Streamchain (streaming transactions one-by-one) is independent of the type of transaction.  }

\textbf{Implications}: {Failures are always reduced regardless of the type of the workload or key distribution.}

\begin{figure*}[!htb]
\centering
\includegraphics[width=0.9\textwidth]{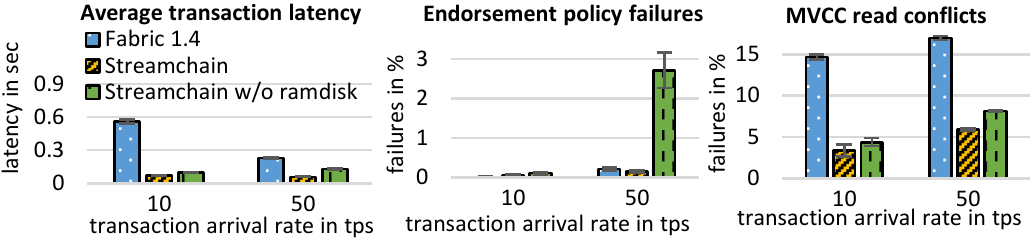}
\caption{Streamchain with and without a RAM disk}
\label{streamnoram}
\end{figure*}

\subsubsection{Effect of RAM disk storage}\label{streamchainram}: Figure~\ref{streamnoram} compares the performance of Streamchain with and without a RAM disk.

\textbf{Observations}: Streamchain with RAM disk performs better than without RAM disk. This is an expected result, as the RAM disk allows for faster reads and writes.
At lower transaction rates, the latency and MVCC read conflicts of Streamchain are improved compared to Fabric, even if there is no RAM disk used. However, at a transaction rate greater than 50 tps, the throughput of Streamchain without RAM disk was too low to sustain the workload, bringing the system into an unstable condition (not shown in the figure). Streamchain cannot handle the streaming of transactions one-by-one without a fast storage at higher transaction rates. 

\textbf{Implications}: The performance improvements of Streamchain are to a large extent caused by the use of a RAM disk storage. The authors of Streamchain have proposed the concept of a virtual block boundary which could be used to commit transactions as blocks while still streaming transactions one-by-one in the ordering service. This concept, if implemented, could potentially remove the need for a RAM disk storage.

\subsection{Results for FabricSharp}
Ruan et al.~\cite{10.1145/3318464.3389693} designed an optimized extension of Fabric which can effectively resolve {MVCC read conflicts}. Similar to Fabric++~\cite{Sharma:2019:BLB:3299869.3319883}, FabricSharp also generates conflict graphs and serializes them. Transactions which cannot be serialized will be aborted before the ordering phase. But unlike Fabric++, FabricSharp generates the conflict graph across blocks and therefore handles both inter-block and intra-block MVCC read conflicts.  

\begin{figure*}[!htb]
\centering
\includegraphics[width=\textwidth]{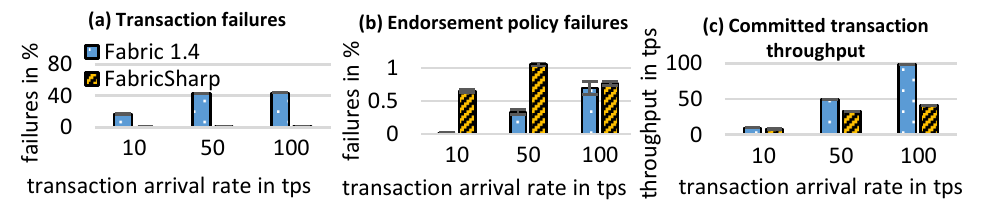}
\caption{Comparison of FabricSharp and Fabric 1.4}
\label{sharp}
\end{figure*}
 
\subsubsection{Transaction failures}: Figure~\ref{sharp} (a) \& (b) compares the performance of FabricSharp and Fabric 1.4 at different arrival rates.

\textbf{Observations}: The transaction failures are significantly lower for FabricSharp. Since all transactions are serialized there are no MVCC read conflicts. FabricSharp does not support range read queries and therefore there are no phantom reads. So only endorsement policy failures are observed for FabricSharp. In FabricSharp, the execution and validation phase are parallelized by using block snapshots at the start of the execution phase. This can introduce stale snapshots that result in more endorsement policy failures.

\textbf{Implications}: FabricSharp is highly effective in resolving MVCC read conflicts but does not resolve endorsement policy failures. 

\subsubsection{Throughput}\label{sharpthroughput}: Figure~\ref{sharp} (c) compares the throughput of FabricSharp and Fabric 1.4 at 10, 50 and 100 tps.

\textbf{Observations}: The committed transaction throughput is lower for FabricSharp. This is an expected result since FabricSharp aborts non-serializable transactions before the ordering phase and only commits successful transactions (and endorsement failures). 

\textbf{Implications}: On the one hand, FabricSharp updates the \linebreak blockchain with only the successful transactions, thus reducing the overhead in the validation phase, but on the other hand, there is no record of failed transactions on the blockchain which could be useful for debugging and auditing purposes.

\begin{figure*}[!htb]
\centering
\includegraphics[width=0.9\textwidth]{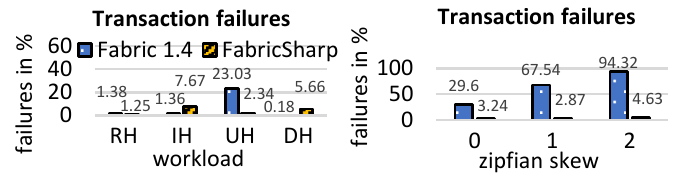}
\captionsetup{labelfont={bf}}
\caption{Effect of workload and skew on FabricSharp}
\label{sharpgen}
\end{figure*}

\subsubsection{{Workloads}}: {Figure~\ref{sharpgen} compares FabricSharp and Fabric 1.4 with different workloads and skew on the C2 cluster with the \emph{genChain} chaincode. We do not use the range-heavy workload because range queries are not supported by FabricSharp.}

\textbf{Observations}: {FabricSharp significantly reduces failures with update-heavy workloads. But FabricSharp does not have a positive effect on insert-heavy and delete-heavy workloads since insert and delete transactions access unique keys which have no dependencies with other transactions. Thus, reordering in FabricSharp can only resolve a limited number of conflicts for these workloads, while the overhead of reordering actually increases the number of failures.}

\textbf{Implications}: {The reordering potential of a workload needs to be analyzed before adopting FabricSharp. This observation is similar for Fabric++ (Section \ref{fabric++sec}). }

\begin{figure*}[!htb]
\centering
\includegraphics[width=0.9\textwidth]{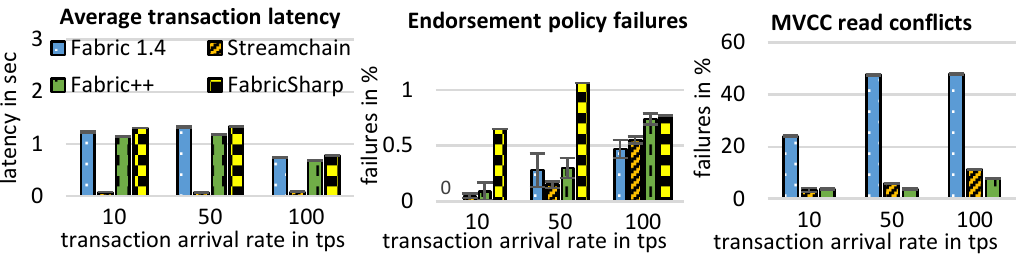}
\caption{Comparison of Fabric systems on C1 cluster}
\label{all4}
\end{figure*}

%\begin{figure}[H]
%%\centering
%\includegraphics[width=\columnwidth]{figures/figure30}
%\captionsetup{labelfont={bf}}
%\caption{Comparison of Fabric systems on C2 cluster}
%\label{all4c2}
%\end{figure}

\subsection{Comparison of Fabric-like systems}

Figure~\ref{all4} compares the latency and transaction failures of all the Fabric-like systems with the EHR chaincode.

\textbf{Observations}: We can observe that Fabric++ and FabricSharp have a similar transaction latency as Fabric 1.4, while Streamchain has a significantly lower latency. We also observe that all three optimizations of Fabric show a significant reduction in the number of failures, but none of them resolve endorsement policy failures. 

\textbf{Implications}: FabricSharp has the best optimization technique to reduce transaction failures when compared to Fabric++ and Streamchain, although it reduces the committed transaction throughput (cf. Section~\ref{sharpthroughput}). The effect of range queries on FabricSharp and Streamchain remains to be studied. Streamchain reduces the latency far better than Fabric++ and FabricSharp, although this is partly due to the use of a RAM disk storage as explained in Section~\ref{streamchainram}. While all three optimizations, Fabric++, Streamchain and FabricSharp, work well in this setting (low transaction arrival rates, no range queries), our previous experiments have revealed some of their limitations.

\section{Lessons Learned}
\label{sec:discussion}
%In this section, we summarize the insights we gained from our experiments and explain with examples how these can be leveraged by a Fabric user. We also discuss future research directions that will be useful to the community at large.
In this section, we summarize the insights we gained from our experiments, explain with examples how these can be leveraged by a Fabric user and discuss future research directions.
\subsection{Insights \& Recommendations}
\textbf{Types of failures}: {The three types of failures described in Section~\ref{sec:failures} are influenced differently by different parameters. Block size has inverse effects on inter-block and intra-block MVCC read conflicts, while it has almost no effect on phantom reads and endorsement policy failures. The number of organizations and endorsement policies have a significant impact on endorsement policy failures while they have insignificant influence on other failures. No parameter tuning in Fabric 1.4 could reduce phantom reads, but reordering the transactions using Fabric++ reduced them. However, Fabric++ could not handle large range reads. FabricSharp early aborts all MVCC read conflicts, but does not resolve endorsement policy failures. Both Fabric++ and FabricSharp increase the number of failures with workloads where reordering possibilities are few, while Streamchain reduces failures regardless of the type of workload. However, Streamchain has a high overhead and lower throughput. Based on our observations, we advice users to analyze their use case, workload and also possibly simulate their network and detect the frequency of different types of failures, before tuning the parameters and using the different Fabric optimizations.
 \\\emph{Example}: If the probability of MVCC read conflicts is high, use Fabric++ or FabricSharp. But if there are very few conflicts these systems will have a negative impact. If phantom reads are high and the range queries have a small range, Fabric++ should be adopted, while for large range queries this will not have an effect. Streamchain should be chosen only if the network expects very low traffic.\\    }   
\textbf{Block size}: The {block size} has a significant effect on the number of transaction failures. The best setting depends on the {transaction arrival rate} and the chaincode. It is a good practice to adapt the {block size} to the arrival rate.\\
\emph{Example}: For the SCM chaincode, we could assume that the holiday seasons would experience high transaction arrival rates if the sales increase. Hence, during those times, change the {block size} to a higher value. Similarly, during off season, decrease the {block size}. \\
\textbf{Number of organizations \& endorsement policies}: Transaction failures increase when the number of organizations and the required number of endorsement signatures increase. Also, sub-policies lead to more failures. So, it is a good practice to lower the number of organizations and create simpler endorsement policies. \\
\emph{Example}: For the SCM chaincode, there will be multiple LSPs, but all of them need not be separate organizations of the Fabric network. LSPs starting from the same source or travelling to the same destination could be grouped together. Also, the LSP organizations need not be a part of the endorsement policy since they are only providing a service. Only the owners and stakeholders need to be part of the endorsement policy. \\
\textbf{Chaincode design \& database type}: LevelDB shows better performance than CouchDB. Users should try to design chaincodes that do not require rich queries, so that LevelDB can be used. Also, since phantom reads are not effectively resolved with any of the Fabric systems, range queries should also be avoided when possible. \\
%\emph{Example}: In the DRM chaincode, there is a \emph{calcRevenue} function which queries the play count of all the music owned by a specific artist. A range query is used to calculate the total revenue of the artist. Instead, every time a song is played, a unique key for each artist to calculate the total revenue could be incremented. This way, a range query could be completely avoided.\\
\emph{Example}: In the DRM chaincode, a range query is used to query the play count of all the music owned by a specific artist and calculate the total revenue. Instead, every time a song is played, a unique key for each artist to calculate the total revenue could be incremented. This way, a range query could be completely avoided.\\
\textbf{Client design}: Read-heavy workloads show lower failures. So, depending on the use case, read-only transactions can be batched at the client side and submitted together. Also, clients can be designed to identify read-only transactions and avoid submitting them to the ordering service since the result of the query is already obtained. Fabric also provides event services~\cite{event} that can be used to update an off-chain database which could be used for read purposes.\\
\emph{Example}: In the SCM chaincode, one needs to read the blockchain multiple times for auditing purposes. It might also be essential to submit these read transactions to the blockchain to keep track of the auditing process. In this scenario, it would be ideal to batch these reads together and submit them when it is not a peak time for other logistic-relating transactions.

\textbf{Our 4 main recommendations for Fabric users are}:
\begin{enumerate}[nosep]
\item Monitor the trend of transaction arrival rates and adapt the block size at appropriate times.
\item Design a Fabric network with fewer organizations, fewer endorsement signatures and fewer endorsement sub-polices. 
\item While designing the chaincode, avoid rich queries and range queries unless they are absolutely necessary.
\item Avoid the submission of read-only transactions to the ordering service or batch them together for submission.
\end{enumerate}
\subsection{Future research directions}
\textbf {Adaptive block size}: A constant block size is not ideal when the transaction arrival rate changes. The ideal block size for various chaincodes is also different. This establishes the need for a dynamically changing block size. Since the transaction arrival rate cannot be determined beforehand and the dependency between arrival rate and block size changes for different chaincodes, it would be useful to monitor the system and adapt the block size dynamically. There are already adaptive blockchain systems that focus on storage or security~\cite{DBLP:journals/corr/abs-1907-13232, 9126003}, but not transaction failures.\\
\textbf {Database optimization}: There is a clear decrease in performance with CouchDB; however, many chaincodes require the use of rich queries. A productive research focus would be to optimize CouchDB or integrate other databases to reduce commit latency in the peers.\\
\textbf {Reduce endorsement policy failures}: Inconsistency of world states is a well-known problem and there is already research in this direction~\cite{li2011determining, yu2007network}. It would be an interesting approach to integrate such research with the Fabric framework and observe the effects on endorsement policy failures.\\
\textbf {Chaincode optimizations}: There is very little research on designing Fabric chaincodes. A challenging research direction would be to analyze different Fabric chaincodes and derive optimization techniques that can reduce transaction failures. 
%This can serve as a guide for future Fabric application developers.

%\textbf {Adaptive block size}: A constant block size is not ideal when the transaction arrival rate changes. The ideal block size for various chaincodes is also different. This establishes the need for a dynamically changing block size. Since the transaction arrival rate cannot be determined beforehand and the dependency between arrival rate and block size changes for different chaincodes, it would be useful to monitor the system and adapt the block size dynamically. \\

\section{Related Work}
\label{sec:related}
%Dinh et al.~\cite{Dinh:2017:BFA:3035918.3064033} proposed the first benchmarking framework for Fabric, Parity and Ethereum. Pongnumkul et al.~\cite{pongnumkul2017performance} also compare Fabric and Ethereum. Both of these approaches are based on Fabric version 0.6 which followed an Order-Execute (O-E) design model based on PBFT consensus. The current version 1.4 of Fabric follows the E-O-V model (cf. Section~\ref{sec:background}) and only supports a crash fault tolerant consensus model. The O-E model and the E-O-V model have significant differences and therefore, the results and observations of these papers are not valid for the current version of Fabric. 

Dinh et al.~\cite{Dinh:2017:BFA:3035918.3064033} and Pongnumkul et al.~\cite{pongnumkul2017performance} present a comparative study of different blockchain frameworks including Fabric but both are based on Fabric version 0.6 which followed an Order-Execute (O-E) design model based on PBFT consensus. The current version of Fabric follows the E-O-V model (cf. Section~\ref{sec:background}) and only supports a crash-fault tolerant consensus model. The O-E model and the E-O-V model have significant differences and therefore, the results of these papers are not valid for the current version of Fabric. 

Many related papers evaluate the performance of Fabric~\cite{8526892, 8525394, Androulaki:2018:HFD:3190508.3190538}. {We go far beyond these existing evaluations and directly compare a large number of systems (Fabric 1.4, Fabric++, Streamchain, FabricSharp) using a large range of different workloads. Further, our focus is on transaction failures, while existing evaluations are mostly concerned with throughput and latency. While throughput and latency are important performance metrics, they are irrelevant if most transactions fail. Similar to our findings, Thakkar et al.~\cite{8526892} also point out the overhead of CouchDB in terms of latency and throughput. In our work, we further explain this overhead by analyzing the latency of each function call in the chaincode. This way, we found that range queries are particularly expensive with CouchDB; a result that has not been reported in~\cite{8526892}.}

%There are also a number of papers that extend Fabric by realizing different optimization techniques but their evaluation focuses only on their implementation and does not extensively compare all the different transaction failures~\cite{Goel:2018:RFP:3284028.3284035, istvan2018streamchain, Sharma:2019:BLB:3299869.3319883, gorenflo2019xox, gorenflo2019fastfabric, 10.1145/3361525.3361540}.
Goel et al.~\cite{Goel:2018:RFP:3284028.3284035} propose a prioritization-based transaction validation model, but they evaluate neither the type nor the number of conflicts. Istv{\'a}n et al.~\cite{istvan2018streamchain} introduce the concept of a virtual block boundary that can reduce the staleness of data used to execute new transactions. However, they do not analyze transaction failures. The main goal of Sharma et al.~\cite{Sharma:2019:BLB:3299869.3319883} with Fabric++ is to reduce the number of transaction conflicts by using optimization strategies of database-like transaction reordering and early aborts. They evaluate two types of MVCC conflicts, but do not discuss endorsement failures and phantom reads. {Further, we identified that the effect of blocksize on transaction failures at a fixed transaction rate is inverse for Fabric 1.4 and Fabric++, which is a new insight. Sharma et al.~\cite{Sharma:2019:BLB:3299869.3319883} only employ a fixed transaction rate and two chaincodes; We employ multiple chaincodes and various transaction rates, so that our analysis is much more comprehensive. Ruan et al.~\cite{10.1145/3318464.3389693} designed an extension of Fabric which also generates conflict graphs and serializes them. However, their evaluation does not show the effect of blocksize on failures, but only on throughput and latency. }

Gorenflo et al.~\cite{gorenflo2019xox} aim to reduce transaction failures by re-executing the conflicting transactions, but they currently have no implementation or evaluation. Gorenflo et al.~\cite{gorenflo2019fastfabric} improve the throughput of Fabric by using multiple optimization strategies. The evaluation is done with a workload of write-only transactions which will never have MVCC read conflicts. Nasirifard et al.~\cite{10.1145/3361525.3361540} use the concept of conflict-free replicated datatypes (CRDT) to resolve conflicts. However, their approach is only applicable for use cases that can be modelled with CRDTs.

{Some of our observations are comparable to research in the database domain. While evaluating SharedDB~\cite{10.14778/2168651.2168654}, a query processing system that batches queries and shares computations, the authors observe an increase in latency with increasing batch size. Similarly, in OLTPShare~\cite{10.14778/3229863.3229866}, a batching scheme for OLTP workloads, smaller batch sizes reduce the potential of sharing while larger batch sizes introduce high latency. Stonebraker et al.~\cite{10.1145/3226595.3226636} observe that developing an application-specific DBMS improves the performance compared to reusing existing DBMS solutions. These three observations are comparable to some of our findings such as that block size in Fabric has a significant influence on failures and that LevelDB, which is embedded in Fabric, performs better than an external database. Though we can draw such parallels with research in the database field, Fabric follows an optimistic concurrency control model that is significantly different from these DBMSs. Additionally, Fabric has other control parameters such as organizations and endorsement policies which are related to blockchains. Thus, the results and inferences in our paper are novel. Also, our work goes beyond existing database research by analyzing the effect of an extensive set of control variables on transaction failures in different extensions of Fabric and focuses on the distributed processing of transactions.}

\section{Conclusions}
\label{sec:conclusion}
In this paper, we formally defined the different transaction failures that occur in Fabric. We designed our own benchmarking system \textsc{HyperLedgerLab} and conducted extensive experiments to analyze the effects of different parameters on failures. We observed a clear dependency between block size and failures, and the optimal block size induced up to 60\% reduction in failures. We also deployed three optimizations of Fabric, Streamchain~\cite{istvan2018streamchain}, Fabric++~\cite{Sharma:2019:BLB:3299869.3319883} and FabricSharp~\cite{10.1145/3318464.3389693}, on \textsc{HyperLedgerLab} and analyzed their performance. We then derived a set of practical recommendations for Fabric users based on our results and discussed possible future research directions. In the future, we will integrate more chaincodes into \textsc{HyperLedgerLab} and also deploy other Fabric optimizations to exhaustively study and compare them.

%\section{Acknowledgments}
\begin{acks}
This work is funded in part by the Deutsche Forschungsgemeinschaft (DFG, German Research Foundation) - 392214008.
%This work is funded by the \grantsponsor{testing} - \grantnum{392214008}
\end{acks}

\newpage

\balance

\bibliographystyle{ACM-Reference-Format}

\bibliography{references} 

\end{document}